\documentclass[11pt,a4paper]{article}

\def\AmSTeX{\leavevmode\hbox{$\mathcal A\kern-.2em\lower.376ex%
        \hbox{$\mathcal M$}\kern-.2em\mathcal S$-\TeX}}
\usepackage{slashed}
\usepackage{color}
\usepackage{makecell}
\usepackage{booktabs}
\usepackage{graphicx}
\usepackage{subfigure}
\usepackage[numbers,sort&compress]{natbib}
\usepackage{amsmath}        
\usepackage{amssymb}
\usepackage{mathrsfs}       
\usepackage{bm}             
\usepackage[amsmath,thmmarks,hyperref]{ntheorem}
\newif\ifpdf \pdftrue
\ifx\pdfoutput\undefined \pdffalse \fi \ifx\pdfoutput\relax
\pdffalse \fi
\ifx\texonly\undefined\let\texonly\relax\fi
\ifx\endtexonly\undefined\let\endtexonly\relax\fi \texonly
  \let\htmlonly\iffalse
  \let\endhtmlonly\fi
\endtexonly

\htmlonly
  
\endhtmlonly
\texonly

  \usepackage{makeidx}

  \usepackage{multicol}
  

\endtexonly
\allowdisplaybreaks
\title{}
\author{\thanks{}}
\date{}
\begin{document}

\title{Lepton-number violating four-body decays of heavy mesons }

\author{Han Yuan\footnote{hanyuan@hit.edu.cn}, Tianhong Wang\footnote{thwang.hit@gmail.com}, Guo-Li Wang\footnote{gl\_wang@hit.edu.cn}, Wan-Li Ju\footnote{wl\_ju\_hit@163.com}\\
{ \it  \small Department of Physics, Harbin Institute of
Technology, Harbin 150001, China}\\Jin-Mei Zhang\footnote{jinmeizhang@tom.com}\\
{\it \small Xiamen Institute of Standardization, Xiamen 361004, China}}
\maketitle

\baselineskip=20pt
\begin{abstract}
Neutrinoless hadron Lepton Number Violating (LNV) decays can be induced by virtual Majorana neutrino, which in turn indubitably show the Majorana nature of neutrinos. Many three-body LNV processes and Lepton Flavour Violating (LFV) processes have been studied extensively in theory and by experiment. As a supplement, we here study 75 four-body LNV (LFV) processes from heavy pseudoscalar $B$ and $D$ decays. Most of these processes have not been studied in theory and searched for in experiment, while they may have sizable decay rates. Since the four-body decay modes have the same vertexes and mixing parameters with three-body cases, so their branching fractions are comparable with the corresponding three-body decays. We calculate their decay widths and branching fractions with current bounds on heavy Majorana neutrino mixing parameters,
and estimate some channels' reconstruction events using the current experimental data from Belle.
\end{abstract}

\section{Introduction}
In the Standard Model (SM), neutrinos are strictly massless,
yet non-zero neutrino masses have been detected in experiment \cite{Eguchi,Ahmed,Argyriades,Barger}. So given the physics of neutrinos, extension of  the SM is necessary. But by now, the nature of neutrinos is still puzzling, because it is still not clear whether neutrinos are Dirac or Majorana particles. So before determine how to extend the physics of SM, we have to clarify the neutrinos type, Dirac or Majorana.

There is strong theoretical motivation for Majorana  mass term to exist since it could naturally explain the smallness of the observed neutrino masses \cite{mink,moha}. As is known, though not derived from first principle, the SM conserves the lepton number, but
Majorana mass term violates lepton number by two units ($\Delta L=2$). In which case the neutrinoless hadron LNV decays with like sign dilepton final state are crucial for the existence of Majorana neutrinos. The possible Lepton Flavor Violating (LFV) meson decays could be induced either by Majorana neutrino or neutrino oscillation in which case the neutrino is a Dirac neutrino. However, neutrino oscillation at loop level would be suppressed by powers of $\frac{m^2_{\nu}}{m^2_{W}}$ and thus the branching fraction could not be brought to an observable level. As a result any direct observation of LFV (LNV) process indicates the existence of Majorana neutrino.

Many efforts have been made to determine the Majorana nature of neutrinos by studying the LNV and LFV processes. As the neutrinos in the final state are undetectable to the detectors, therefore neutrinoless processes are preferred, {\it e.g.}, the neutrinoless double $\beta$ nuclei decay ($0\nu\beta\beta$) has long been advocated as a premier demonstration of possible Majorana nature of neutrinos \cite{klei,Avigone}; the Majorana neutrino exchanges in $\tau$ lepton three-body or four-body decays \cite{castro-1,castro-3,chrzqszcz}; the LNV process $pp\rightarrow\ell^{\pm}\ell^{\pm}+X$ or $pp\rightarrow\ell^{\pm}\ell^{\pm}jj$ at LHC \cite{W.Chao1,W.Chao2}; the top-quark or W-boson four-body decay \cite{Bar,Z.G.}; the LNV or LFV meson decays with  like sign dilepton  in the final state \cite{dib,ali,Mikhail,Atre2,han-1,kim-1}, {\it et al}.

Recently, Atre et al. \cite{Atre1} have studied $K$, $D$, $D_s$ and $B$ decays via a fourth massive Majorana neutrino. They demonstrated if the exchanged Majorana neutrino is resonant, which means it is on mass-shell. Then the corresponding branching fractions can be enhanced by several orders, in which case the fractions can be reached by the current experiments. Inspired by the effect of resonant neutrino, various three body meson decays $M_1^+\rightarrow\ell_1^+\ell_2^+M_2^-$ where $\Delta L=2$ have been studied in \cite{Laurence1,Laurence2,Atre1,Cvetic,Jin,chen-1} and so have four-body decays ${B}\rightarrow D\ell\ell\pi$ in \cite{David}.

In the experiment, some of these LNV (LFV) processes have been searched. For example,
Fermilab E791 Collaboration reported their results of searching for the LNV and LFV decays of $D^0$ into 3 and 4-bodies, they presented  upper limits on the branching fractions at $90\%$ confidence level (CL) \cite{Aitala}. Recently, using $772\times10^6$ $B\bar B$ pairs accumulated at $\Upsilon(4S)$ resonance with the same CL, the Belle Collaboration set the upper limits on the LNV (LFV) $B^+ \to D^- \ell^+ \ell^{'+}$ decays \cite{belle-1}. Using a sample of $471\pm 3$ million $B\bar B$ events, the BABAR Collaboration searched for the LNV processes $B \to K^-(\pi^-) \ell^+ \ell^+$ and placed upper limits on their branching fractions also with $90\%$ CL \cite{babar-1}. The LHCb Collaboration, using $0.41$ fb$^{-1}$ of data collected with the LHCb detector in proton-proton collisions at a center-of-mass energy of $7$ TeV, reported their upper limits on the branching fractions of $B^-$ decays to $D^{(*)+}\mu^-\mu^-$, $\pi^+\mu^-\mu^-$ and $D_s\mu^-\mu^-$ at $95\%$ CL. They also searched for the 4-body decay $B^-\rightarrow D^0\pi^+\mu^-\mu^-$ and set upper limit on its branching fraction \cite{Aaij} for the first time. The experimental situation of searching for the LNV and LFV processes can be found in Refs.~\cite{Seyfert,chrzqszcz}. Though these LNV and LFV processes are still unobservable, the upper limits for branching fraction have been obtained, which also in turn limit the mixing parameters between Majorana neutrino and charged lepton.

Though lots of LNV (LFV) processes have been studied by experiment and in theory, there are still many channels which have not been considered, especially the four-body LNV (LFV) meson decays, most of which are still absent in literature. Some channels of that kind may have considerable branching fractions and may be accessible in current experiment. LNV four-body decays also offer complementary information about the masses and heavy mixings of such a heavy (resonant) Majorana neutrino, so they are worth studying deeply. In this paper, we study 75 four-body LNV (LFV) processes of dilepton decays $B(D)\rightarrow M_1\ell_1\ell_2M_2$, where $M_1$ stands for a pseudoscalar meson, $M_2$ can be a pseudoscalar or a vector meson and $\ell_1(\ell_2)=e,~\mu$. These $\Delta L=2$ LNV (LFV) 4-body meson decays are induced by a Majorana neutrino, and the possible lowest order diagrams are illustrated in Figure~1~(a-b). Some processes, such as the decays of ${\bar B}^{0}\rightarrow D^- \ell^+_1 \ell^+_2 M^{-}_2$, where $M^{-}_2$ stands for $\pi^-$, $K^-$, $\rho^-$, $K^{*-}$, $D^-$ or $D^{-}_s$, are represented by an exclusive Feynmann diagram shown in Figure~1~(a); but some decays, like ${B^+}\rightarrow \bar {D}^0 \ell^+_1 \ell^+_2 M^{'-}$, where $M^{'-}$ denotes $D^{-}$ or $D^{-}_s$, have both decay modes shown in Figure~1~(a) and (b).
In Figure~1~(a), if the Majorana neutrino mass lies between a few hundred MeV to 4.4 GeV (since it is heavy, it may be a fourth generation neutrino), the neutrino could be on mass-shell (resonance), and the corresponding decay rate will be much enhanced due to the effect of neutrino-resonance. The contribution of Figure~1~(a) will be much greater than that of neutrino-exchange diagram in Figure~1~(b), which is suitable for a continuous neutrino mass. So we will focus on the neutrino-resonance of diagram figure 1~(a). The contribution of neutrino-exchange diagram figure 1~(b) and the interference between two diagrams will be ignored.

There are two key points to calculate these 4-body decay modes. One is the selection of the mixing parameters, since most of these Majorana neutrino induced 4-body decay modes do not have experiment results and we cannot extract the mixing parameters by these decays. So we followed Atre {\it et al}'s method in which the parameters are determined by experimental data \cite{Atre1}. We choose the strongest constrains which were abstracted from the current data as the input in our paper  \cite{Atre1,Jin} to guarantee the accuracy. The other point is the calculation of the hadronic matrix element between initial meson $B(D)$ and final meson $M_1$. We use the Mandelstam formalism \cite{mandel} which the hadronic matrix element is described as an overlapping integral over the wave functions of the initial and final states \cite{gauge}. The wave functions are obtained by solving the relativistic Bethe-Salpeter (BS) equation \cite{BS}.

This paper is organized as followed, in section 2, we outline the formulas of the transition matrix element. In section 3, we present the details of how to calculate the hadronic matrix element. In section 4, we show the results and conclude the branching fraction of heavy meson 4-body decays as a function of the heavy neutrino mass.

\section{Theoretical Details}

The leading order Feynman diagrams for the LNV (LFV) 4-body decays of heavy meson $M$:
\begin{equation}\label{eq:decayprogram}
M(P){\rightarrow}M_1(P_1){\ell}^{+}_1(P_2){\ell}^{+}_2(P_3)M^{-}_2(P_4)
\end{equation}
are shown in Figure~1~(a-b). Here $M$ is the pseudoscalar $B$ or $D$ with momentum $P$,  two charged leptons $\ell^+_1$, $\ell^+_2$ have momentum $P_2$ and $P_3$, pseudoscalar meson $M_1$ with momentum $P_1$ denotes $\pi$, $K$ or $D$, meson $M_2$ with momentum $P_4$ can be a pseudoscalar meson $\pi$, $K$, $D$ and $D_s${\it et al}, or a vector meson $\rho$, $K^*$, {\it et al}.

Such LNV (LFV) process can occur through a Majorana neutrino, and the vertex between this Majorana neutrino and charged lepton is beyond the SM. Following previous studies \cite{Atre1,Bar-Shalom}, we assumed that there is only one heavy Majorana neutrino which may be a fourth generation neutrino. It can be kinematically accessible in the range we are interested in. Then the gauge interaction lagrangian responsible for the LNV (LFV) decay can be written as:
\begin{equation}
\mathcal{L}=-\frac{\mathit{g}}{\sqrt{2}}W^+_\mu \sum\limits^{\tau}\limits_{\ell=e}V^*_{\ell 4}\overline{N^c_{4}}\gamma^{\mu}P_L\ell+\mathrm{h.c.},
\end{equation}
where $P_L=\frac{1}{2}(1-\gamma_5)$, $N_{4}$ is the mass eigenstate of the fourth generation Majorana neutrino and $V_{\ell 4}$ is the mixing matrix between the charged lepton $\ell$ and heavy Majorana neutrino $N_4$.

\begin{figure}[!h]\label{fig:feyn}
\centering
\subfigure {}
\addtocounter{subfigure}{-2}
\subfigure[]
{\subfigure{\includegraphics[scale=0.6]{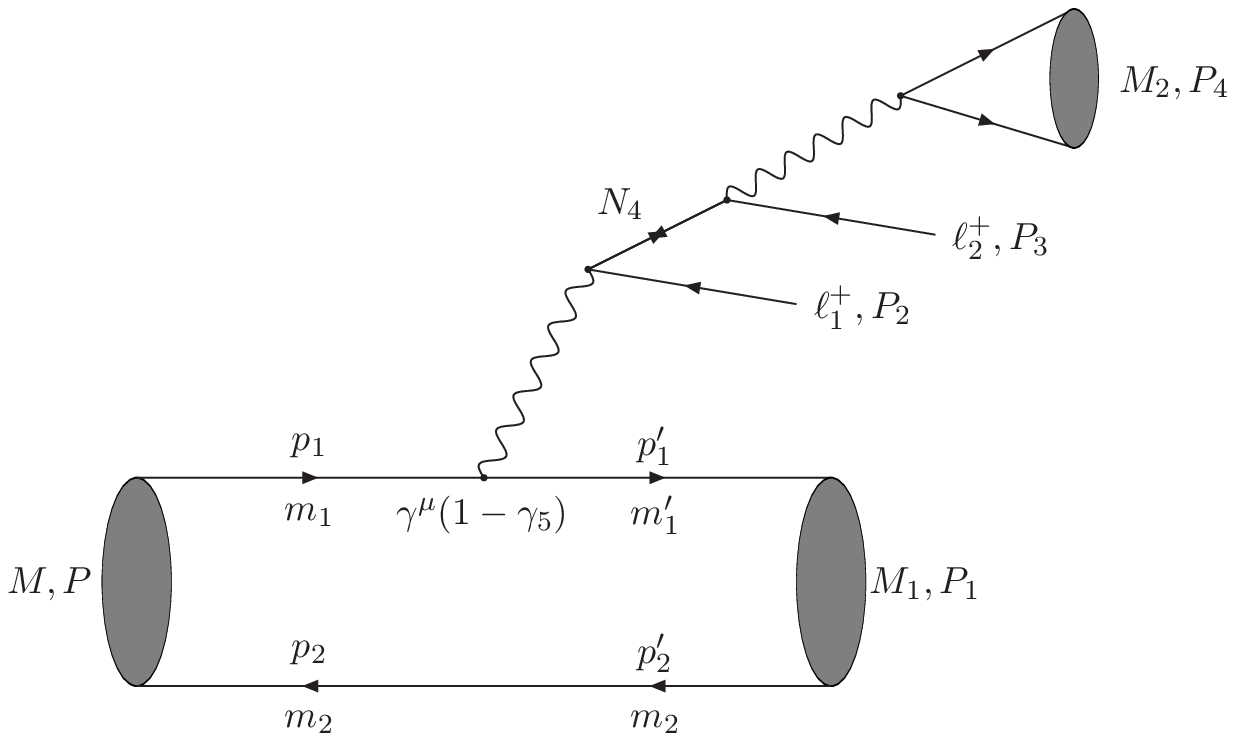}}}
\subfigure {}
\addtocounter{subfigure}{-2}
\subfigure[]
{\subfigure{\includegraphics[scale=0.5]{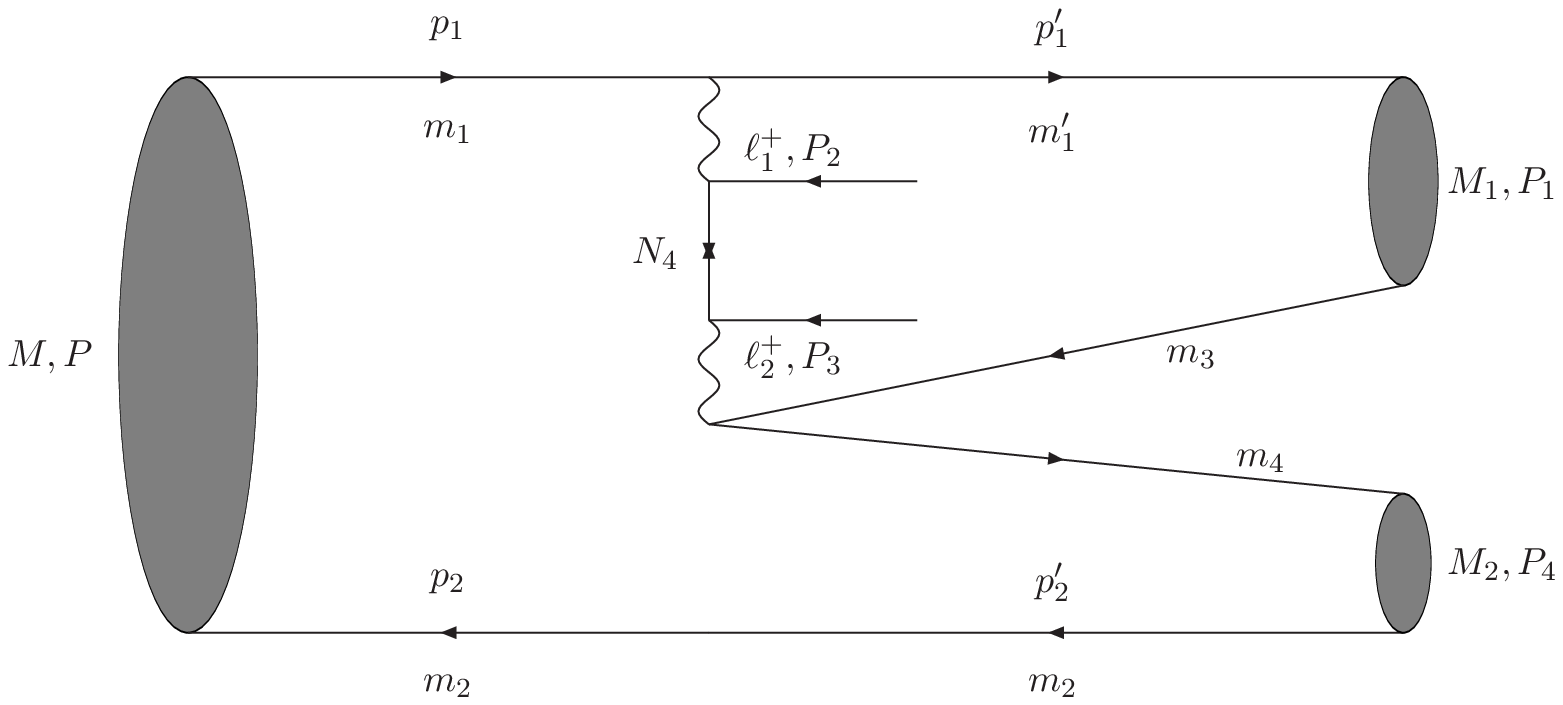}}}
\caption{Feynman diagram of the four-body decay of heavy meson}\vspace{-1em}
\end{figure}

The transition amplitude for the 4-body decay $M(P){\rightarrow}M_1(P_1){\ell}^{+}_1(P_2){\ell}^{+}_2(P_3)M^{-}_2(P_4)$ shown in Figure~1~(a) can be written as:
\begin{equation}\label{eq:firstam}
\mathcal{M}=\frac{g^2 V_{q_1q_2} V_{q_3q_4}}{8M_W^4}\langle M_1(P_1)|{\bar{q}_1}\gamma^\mu(1-\gamma_5)q_2 |M(P)\rangle\times \mathcal{M}_{\mu\nu}\times \langle M_2(P_4)|{\bar{q}_3}\gamma^\nu(1-\gamma_5)q_4|0\rangle,
\end{equation}
where the momentum dependence in the propagator of $W$ boson has been ignored since it is much smaller than the $W$ mass; $g$ is the weak coupling constant; $V_{q_1q_2}$ ($V_{q_3q_4}$) is  the Cabibbo-Kobayashi-Maskawa (CKM) matrix element between quarks $q_1$ and $q_2$ ($q_3$ and $q_4$); $\mathcal{M}_{\mu\nu}$ is the transition amplitude of the leptonic part.

As mentioned before, only the contribution of the diagram in Figure~1~(a) is considered, where the Majorana neutrino is on mass shell, and the effective narrow-width approximation related to the resonant contribution can enhance the decay rate substantially. In this case,
according to Ref.~\cite{Atre1,Jin}, the leptonic matrix element $\mathcal{M}_{\mu\nu}$ can be given as:
\begin{equation}\label{eq:DeltaL}
\mathcal{M_{\mu\nu}}=\frac{\mathit g^2}{2}V_{\ell_14}V_{\ell_24}m_4\left[\frac{{\bar{u}_1}\gamma_{\mu}\gamma_{\nu} P_R\nu_2}{q_{{_N}_4}^2-m_4^2+i\Gamma_{N_4}m_4}+\frac{{\bar{u}_1}\gamma_{\nu}\gamma_{\mu} P_R\nu_2}{q^{\prime2}_{{_N}_4}-m_4^2+i\Gamma_{N_4}m_4}\right],
\end{equation}
where $V_{\ell4}$ is the mixing parameter between the heavy Majorana neutrino and charged lepton, $P_R=\frac{1}{2}(1+\gamma_5)$; $q_{{_N}_4}$ is the momentum of heavy Majorana neutrino ($q^{\prime}_{{_N}_4}$ is the case of exchange the two final charged leptons), $m_4$ is the mass of the heavy Majorana neutrino and $\Gamma_{N_4}$ is the total decay width of the heavy neutrino.

Mesons $M$ and $M_1$ are pseudoscalar mesons and the corresponding hadronic matrix element in Eq. (\ref{eq:firstam}) can be described as a function of form factors:
\begin{equation}\label{eq:hadronic}
\langle M_1(P_1)|{\bar{q}_1}\gamma^\mu(1-\gamma_5)q_2|M(P)\rangle=P^\mu(f_++f_-)+P^\mu_1(f_+-f_-),
\end{equation}
The method to calculate the form factors $f_+$, $f_-$ will be shown in section 3.

The last part $\langle M_2|h_2^\nu|0\rangle$ in Eq.~(\ref{eq:firstam}) is related to the decay constant of the meson $M_2$. If $M_2$ is a pseudoscalar with momentum $P_4$, we obtain the following relation:
\begin{equation}\label{eq:constant}
\langle M_2(P_4)|{\bar{q}_3}\gamma^\nu(1-\gamma_5)q_4|0\rangle=i F_{M_2}P_4^\nu,
\end{equation}
where $F_{M_2}$ is decay constant of meson $M_2$. If $M_2$ is a vector with momentum $P_4$ and polarization vector $\epsilon$, the corresponding relation will become:
\begin{equation}\label{eq:constant1}
\langle M_2(P_4,\epsilon)|{\bar{q}_3}\gamma^\nu(1-\gamma_5)q_4|0\rangle=M_2F_{M_2}{\epsilon}^\nu,
\end{equation}here we use the same symbol $M_2$ to denote the meson and its mass.

By combining Eq.~(\ref{eq:DeltaL}), Eq.~(\ref{eq:hadronic}) and Eq.~(\ref{eq:constant}), we rewrite the decay amplitude Eq.~(\ref{eq:firstam}) in the case of meson $M_2$ as a pseudoscalar:
\begin{eqnarray}\label{eq:amplitude}
\mathcal{M}&=&2G_F^2V_{\ell_{1}4}V_{\ell_{2}4}V_{q_1q_2}V_{q_3q_4} F_{M_2}m_4\nonumber\\
&&{\times}{\bar{u}_1}\left[{{\frac{{\not\!P}{\not\!P_4(f_++f_-)}+{\not\!P_1}{\not\! P_4(f_+-f_-)}}{(P_3+P_4)^2-m^2_4+i{\Gamma}_{N_4}m_4}}{+{\frac{{\not\!P_4}{\not\!P(f_++f_-)}+{\not\!P_4}{\not\! P_1(f_+-f_-)}}{(P_2+P_4)^2-m^2_4+i{\Gamma}_{N_4}m_4}}}}\right]P_R\nu_2,
\end{eqnarray}
where $G_F$ is Fermi constant. If meson $M_2$ is a vector, we just replace ${\not\!P_4}$ with $M_2{\not\!\epsilon}$ in numerator in Eq.~(\ref{eq:amplitude}). With the numerical values of form factors $f_+$ and $f_-$ obtained in section 3, the calculation of this decay amplitude is not complicated.

\section{Hadronic transition matrix element}

In order to calculate the hadronic matrix element and get the numerical value of form factors $f_+$, $f_-$, we use the Mandelstam formalism \cite{mandel}, in which the transition amplitude between two mesons is described as a overlapping integral over the Bethe-Salpeter wave functions of initial and final mesons \cite{gauge}. Using this method with further instantaneous approximation \cite{salpe}, in the center of mass system of initial meson, in leading order, we write the hadronic matrix element as \cite{fu}:
\begin{equation}\label{eq:amjuti}
\langle M_1(P_1)|{\bar{q}_1}\gamma^\mu(1-\gamma_5)q_2|M(P)\rangle=\int\frac{\mathrm{d}\vec{q}}{(2\pi)^3}\mathrm{Tr}
\left[\bar{\varphi}_{P_1}^{++}(\vec{q}_1)\gamma_\mu(1-\gamma_5)\varphi_P^{++}(\vec{q})\frac{\not\!P}{M}\right],
\end{equation}
where $P$ and $P_1$ are the momenta of initial and final mesons; $M$ in denominator is the mass of initial meson; $q$ is relative momentum between quark and antiquark inside the initial meson; $\vec{q}_1=\vec{q}+\frac{m_{2}}{m^{\prime}_{1}+ma^{\prime}_{2}}\vec{r}$ is the relative momentum inside the final meson $M_1$, $m^{\prime}_{1}$ ($m_{2}^{\prime}$) is mass of antiquark (quark) in final meson $M_1$, $\vec{r}$ is three dimension momentum of meson $M_1$; $\varphi^{++}$ is the positive wave function for a meson in the BS method; for the final state, we have define the symbol $\bar{\varphi}_{P_1}^{++}=\gamma_0(\varphi_{1P_1}^{++})^{+}\gamma_0$.
\begin{table}[htbp]
\caption{ Mass of quark in unit of GeV. }
\vspace{0.5em}\centering
\begin{tabular}{cccccc}
\toprule[1.5pt]
 quark & $b$ & $c$ & $s$ & $d$ & $u$\\
\midrule[1pt]
 mass&$4.96$&$1.62$&$0.5$&$0.311$&$0.305$\\
\bottomrule[1.5pt]
\end{tabular} \label{quark}
\end{table}

In the BS method, the positive wave function $\varphi^{++}$ for a pseudoscalar meson can be written as \cite{me0-}:
\begin{equation}\label{eq:wave}
\varphi^{++}_{P}=A\left(B+\frac{\not\!P}{M}+{\not\!q}_{\perp}C+\frac{{\not\!q}_{\perp}{\not\!P}}{M}D\right)\gamma_5,
\end{equation}
where $q_{\perp}=(0,\vec q)$, and
\begin{eqnarray}\label{eq:ABCD}
A&=&\frac{M}{2}\left[f_1(\vec{q})+f_2(\vec{q})\frac{m_1+m_2}{\omega_1+\omega_2}\right],\nonumber\\
B&=&\frac{\omega_1+\omega_2}{m_1+m_2},\nonumber\\
C&=&-\frac{m_1-m_2}{m_1\omega_2+m_2\omega_1},\\
D&=&\frac{\omega_1+\omega_2}{m_1\omega_2+m_2\omega_1}.\nonumber
\end{eqnarray}
In Eq.~(\ref{eq:ABCD}), $m_1$ and $m_2$ are the masses of quark and antiquark inside the meson, and we list their values in Table~\ref{quark};
$\omega_i$ is defined as
$\omega_i=\sqrt{m_i^2+\vec q^2}$, $i=1,2$; $f_1(\vec{q})$ and $f_2(\vec{q})$ are the wave function of the meson.

With Eq.~(\ref{eq:wave}) and Eq.~(\ref{eq:ABCD}), we take the integral on the right side of Eq.~(\ref{eq:amjuti}), then the form factor $f_+$, $f_-$ can be expressed as:
\begin{eqnarray}
f_+=\frac{1}{2}\left(\frac{T_1}{M}+\frac{T_2}{M_1}+\frac{M-E_1}{M}T_3\right),\nonumber\\
f_-=\frac{1}{2}\left(\frac{T_1}{M}-\frac{T_2}{M_1}-\frac{M+E_1}{M}T_3\right),
\end{eqnarray}
where $M_1$ and $E_1=\sqrt{M_1^2+\vec r^2}$ are the mass and energy of final meson $M_1$; and
\begin{eqnarray}\label{eq:ttt}
T_1&=&\int\frac{d^3\vec{q}}{(2\pi)^3}4A_1At_1,\nonumber\\
T_2&=&\int\frac{d^3\vec{q}}{(2\pi)^3}4A_1At_2,\nonumber\\
T_3&=&\frac{1}{|\vec{r}|}\int\frac{d^3\vec{q}}{(2\pi)^3}4A_1At_3|\vec{q}|\cos\theta,\nonumber\\
t_1&=&C_1\frac{m_{12}}{m_{11}+m_{12}}E_1-\frac{C\vec{q}_1\cdot\vec{q}}{M_1}+\frac{BD_1}{M_1}\left(\vec{q}_1\cdot\vec{q}+
\frac{m_{12}}{m_{11}+m_{12}}\vec{q}_1^2+\frac{m_{12}}{m_{11}+m_{12}}E_1^2\right)\nonumber\\
&&-CC_1\left(\vec{q}^2+\frac{m_{12}}{m_{11}+m_{12}}\vec{q}_1\cdot\vec{q}\right)
-BB_1-DD_1\frac{m_{12}}{m_{11}+m_{12}}\frac{E_1}{M_1}\vec{q}_1\cdot\vec{q},\nonumber\\
t_2&=&-1-\frac{m_{12}}{m_{11}+m_{12}}C_1M_1-\frac{m_{12}}{m_{11}+m_{12}}BD_1E_1-DD_1\vec{q}^2,\nonumber\\
t_3&=&-C_1-B_1D-BD_1\frac{E_1}{M_1}-\frac{CE_1}{M_1}+\frac{DD_1}{M_1}
\left(2\vec{q}_1\cdot\vec{q}+\frac{m_{12}}{m_{11}+m_{12}}\vec{q}_1^2\right),
\end{eqnarray}
where $A_1$, $B_1$, $C_1$ and $D_1$ have the same meanings as those in Eq.~(\ref{eq:ABCD}), while the parameters are replaced by the one of final pseudoscalar.

Numerical values of wave functions $f_1(\vec{q})$ and $f_2(\vec{q})$ can be obtained by solving the coupled Salpeter equations \cite{me0-}:
\begin{eqnarray}\label{eq:ouhe}
(M-2\omega_1)\left[f_1(\vec{q})+f_2(\vec{q})\frac{m_1}{\omega_1}\right]
&=&-\int\frac{d\vec{k}}{(2\pi)^3}\frac{1}{\omega_1^2}\left\{(V_s-V_v)\left[f_1(\vec{k})m_1^2\right.\right.\nonumber\\
&&\left.\left.+f_2(\vec{k})m_1\omega_1\right]-(V_s+V_v)f_1(\vec{k})\vec{k}\cdot\vec{q}\right\},\nonumber\\
(M+2\omega_1)\left[f_1(\vec{q})-f_2(\vec{q})\frac{m_1}{\omega_1}\right]
&=&-\int\frac{d\vec{k}}{(2\pi)^3}\frac{1}{\omega_1^2}\left\{(V_s-V_v)\left[f_1(\vec{k})m_1^2\right.\right.\nonumber\\
&&\left.\left.-f_2(\vec{k})m_1\omega_1\right]-(V_s+V_v)f_1(\vec{k})\vec{k}\cdot\vec{q}\right\}.
\end{eqnarray}
where we have chosen the Cornell potential, which is a linear potential plus a single gluon exchange reduced vector potential, and in momentum space the expression is:
\begin{eqnarray}
V_s(\vec{q})&=&-\left(\frac{\lambda}{\alpha}+V_0\right)\delta^3(\vec{q})+\frac{\lambda}{\pi^2}\frac{1}{(\vec{q}^2+\alpha^2)^2},\nonumber\\
V_v(\vec{q})&=&-\frac{2}{3\pi^2}\frac{\alpha_s(\vec{q})}{(\vec{q}^2+\alpha^2)},\nonumber\\
\alpha_s(\vec{q})&=&\frac{12\pi}{27}\frac{1}{\log(a+\frac{\vec{q}^2}{\Lambda_{QCD}^2})},
\end{eqnarray}
where $a=e=2.71828$; $\lambda=0.21$ GeV$^2$ is the string constant; $\alpha=0.06$ GeV is a parameter for the infrared divergence compensation; the QCD scale $\Lambda_{QCD}=0.27$ GeV characterizes the running strong coupling constant $\alpha_s$; the constant $V_0$ is a parameter by hand in potential model to match the experimental data, whose values for different mesons are listed in Table.~{\ref{vv0}}.

\begin{table}[htbp]
\caption{ Parameters  $V_0$ in unit of GeV }
\vspace{0.5em}\centering
\begin{tabular}{ccccc} \toprule[1.5pt]
meson &$B$&$D$&$K$&$\pi$ \\\midrule[1pt]
$V_0$&-0.091&-0.375&-0.962&-0.999\\\bottomrule[1.5pt]
\end{tabular} \label{vv0}
\end{table}

With these parameters, we solved the full Salpeter equation Eq.~(\ref{eq:ouhe}), and obtained the numerical values of wave functions $f_1(\vec{q})$ and $f_2(\vec{q})$ for pseudoscalar mesons $B$, $D$, $K$ and $\pi$. Meanwhile, the meson masses of these pseudoscalar mesons are also obtained which agree with experimental data.

\section{Numerical Results and Discussions}

Besides the parameters appearing in potential, there are other parameters whose values need to be determined. We choose the CKM matrix elements \cite{PDG}: $V_{ud}=0.974$, $V_{us}=0.225$, $V_{cd}=0.230$, $V_{cs}=0.973$, $V_{cb}=40.6\times10^{-3}$, $V_{ub}=3.89\times10^{-3}$.
The decay constants of pseudoscalar and vector mesons used in our calculation are listed in Table~\ref{tab3}.
\begin{table}[htbp]
\caption{ Decay constants $F_{M_2}$ of pseudoscalar and vector mesons in unit of MeV. }
\vspace{0.5em}\centering
\begin{tabular}{ccccccc} \toprule[1.5pt]
meson&$\pi$&$\rho$&$K$&$K^{*}$&$D$&$D_s$\\\midrule[1pt]
$F_{M_2}$&130.4~\cite{PDG}&220~\cite{Ebert}&156.1~\cite{PDG}&217~\cite{Ebert}&222.6~\cite{Artuso}&260~\cite{PDG}\\\bottomrule[1.5pt]
\end{tabular} \label{tab3}
\end{table}

The key step to calculate the decay widths and branching fractions of LNV (LFV) heavy meson decays is to determine the limits on the mixing parameters $|V_{\ell_14}V_{\ell_24}|$ and the heavy neutrino mass $m_4$ in Eq.~(\ref{eq:DeltaL}). Following the approaches in Refs.~\cite{Atre1,Jin}, we take the mixing parameter $V_{\ell4}$ and the mass $m_4$ as phenomenological parameters. Since the mixing parameters are common constant, we take some decay modes with the same $|V_{\ell_14}V_{\ell_24}|$ into our consideration and have mixing parameters numerical upper bounds in experiment, thus we extract the numerical values of mixing parameters from these processes. Details can be found in Refs.~\cite{Atre1,Jin}. We choose the strongest constrains on mixing as input in this paper to guarantee accuracy.
For the value of $m_4$, since we only consider the case in which the heavy neutrino is on mass shell, we determine the mass of neutrino by kinematics. With numerical values of mixing parameters and neutrino mass $m_4$, the neutrino total decay width $\Gamma_{N_4}$ is calculated, which covers all possible decay channels of Majorana neutrino at the mass $m_4$ \cite{Atre1}. So in our calculation, $\Gamma_{N_4}$ is not fixed but mass and mixing parameter dependent.

With these parameters and the limits on mixing parameters, 75 LNV (LFV) decay widths and branching fractions of the heavy mesons $D^+$, $D^0$, $B^+$, and $B^0$ are calculated.
Among these processes, there are some channels where the meson $M_1$ is a light meson, $\pi$ or $K$. We must point out that since we have made instantaneous approximation to
Bethe-Salpeter equation, the result of the hadronic matrix element including a light meson may not be accurate in the heavy meson case. Since all these decays are beyond the SM, accurate calculation is not the issue, and we also take the results including these decays. In the calculation of decay rate, we perform a Monte Carlo sampling of the branching fractions and the mass of heavy neutrino. For example we calculate the excluded region of the branching fractions as a function of the heavy neutrino mass $m_4$ and plot the results in Figures~2-7. The regions inside and above the curve are excluded by current experiment data, while the region below the curve is allowed in theory.

The curve is not smooth, which is caused by two reasons. First, we choose different mixing parameters $|V_{\ell_14}V_{\ell_24}|$ according to different ranges of heavy neutrino mass $m_4$. Since the current limits on mixing parameters are related to heavy Majorana neutrino mass, depending on to different neutrino mass range, we choose different LNV (LFV) processes to get the strongest constrains on mixing parameters. For example, in process $B^0{\rightarrow}D^-e^+e^+M_2^-$, we choose three processes $K^+\rightarrow e^+e^+\pi^-$, $D^+\rightarrow e^+e^+\pi^-$ and $B^+\rightarrow e^+e^+\pi^-$ to limit $|V_{e4}|^2$. Second, as discussed above, the value of neutrino total decay width $\Gamma_{N_4}$ is mass and mixing parameter dependent, whose values change with respect mixing parameter $|V_{\ell 4}|$ and neutrino mass $m_4$.  Because the mixing parameters is piecewise, the branching fractions are also piecewise as a function of the neutrino mass. The difference of value choices of mixing parameters may be the main reason for the difference between our results and those in Ref.~\cite{David,castro-2}, which calculated the branching fractions of $\bar{B}^0\rightarrow D^+ e^-e^-\pi^+$ and ${B}^-\rightarrow D^0 \mu^-\mu^-\pi^+$. We mention that, in calculations of the decay modes $B^+\rightarrow \pi^0\ell_1^+\ell_2^+M_2^-$, we lack the information of mixing parameter $|V_{e4}V_{\mu4}|$ when neutrino mass $m_4>4~\mathrm{GeV}$, so the results in Figure~6~(b) are given by set $|V_{e4}V_{\mu4}|=0$ in these cases, that is, there are no predictions when neutrino mass $m_4$ is larger than $4~\mathrm{GeV}$ in Figure~6~(b).

There is another point that seems unusual in the results of some branching fractions. For example in Fig.~5~(b), we show the branching fraction for the decay mode $D^+\rightarrow \bar K^0e^+\mu^+K^-$. At two edges of the curve, which are the points of the allowed smallest and largest neutrino masses separately, the values of the branching fractions are very small. The small rates is not unusual actually, because it happens due to the restriction of the phase space. The very small kinematic phase space at edges lead to those small branching fractions.

Because some 4-body decays of mesons have broader phase space than the corresponding 3-body processes and the resonance neutrino mass is determined kinematically, one of the advantages of these 4-body decays is that we can detect much wider range of Majorana neutrino mass. For example, we can study the heavy neutrino if its mass is in the range of 2 GeV $\thicksim$ 4 GeV durning the 3-body decay $B^-\rightarrow e^-e^-D^+$ \cite{Jin}. While durning the 4-body decay $B^0\rightarrow D^-e^+e^+\pi^-$ or $B^+\rightarrow \bar D^0e^+e^+\pi^-$, we can reach the range of possible neutrino mass from 0.2 GeV to 3.4 GeV. Another advantage is that the branching fraction is not small compared with the corresponding 3-body decay \cite{Atre1,Jin}. Because, in some cases, they have same vertexes, mixing parameters $|V_{\ell_14}V_{\ell_24}|$ and CKM matrix elements.

We have mentioned that the dominant factors of the branching fractions comes from the mixing parameter $|V_{\ell_14}V_{\ell_24}|$, which are limited by the current experimental data. Besides these parameters, there are other important parameters: CKM matrix elements, which are also determinant factors to the values of branching fractions. We note that if the final mesons are $D^-$ and $\pi^-$, there are two decay modes, $B^0\rightarrow D^-\ell_1^+\ell_2^+\pi^-$ and $B^0\rightarrow \pi^-\ell_1^+\ell_2^+D^-$. In the first decay mode, the CKM matrix elements are $|V_{cb}V_{ud}|^2$, while for the second are $|V_{ub}V_{cd}|^2$, as $|V_{ub}V_{cd}|^2/|V_{cb}V_{ud}|^2\sim 4\times 10^{-4}$, so we ignore the decay $B^0\rightarrow \pi^-\ell_1^+\ell_2^+D^-$ and its interference with $B^0\rightarrow D^-\ell_1^+\ell_2^+\pi^-$. For the same reason we only consider the contribution of decay $D^0\rightarrow K^-\ell_1^+\ell_2^+\pi^-$ and ignore the decay mode $D^0\rightarrow \pi^-\ell_1^+\ell_2^+K^-$.

In some particular channels, there is an additional contribution coming from intermediate mesons resonance \cite{castro-2}. For example, in the decay channel $B^{+}{\rightarrow}{\bar D}^0\mu^+\mu^+\pi^-$, besides the CKM favored diagram in Figure~1~(a), there is another CKM dis-favored diagram (see Figure~1~(b) in Ref.~\cite{castro-2}), where the two final mesons can be induced by a intermediate resonance $D^{*-}(2010)$, in range of $2.1~\mathrm{GeV}\leq m_4\leq 3.3~\mathrm{GeV}$. And the intermediate resonance $D^{*-}(2010)$ may results in a considerable contribution in decay $B^{+}{\rightarrow}{\bar D}^0\mu^+\mu^+\pi^-$, but we do not take into consideration these cases.

Some channels with large branching ratios are detectable by the current experiments. For example, the Belle Collaboration produced $772$ million $B\bar{B}$ events per year \cite{Belle1}, which can be used to study the four-body $B$ meson LNV and LFV decays. For Belle detector, the reconstruction efficiencies of $\pi^0$, $\rho$, $K^*$, $D$, $D^0$ and $D_s$ are $65\%$, $61\%$, $58\%$, $78\%$, $83\%$ and $74\%$, respectively; the identification efficiencies of $\pi^\pm$ and $K^\pm$ are $95\%$ and $86\%$ \cite{Belle2}; the electrons and muons efficiency rates both approximate $90\%$ \cite{Belle1}. With these  efficiencies, we choose maximum branching fractions interval in each process, and estimate the reconstruction events shown in Table~\ref{B0D}, Table~\ref{BD0} and Table~\ref{Bpi}. Of particular note, all the results do not include the
influence of Geometrical Acceptance. The reason why we do not calculate $K$ and $D$ reconstruction events in Table \ref{Bpi}, is that the branching fractions of $K$ and $D$ are too small, which is less than the $B\bar{B}$ events.

There are 3 million $D^0\bar{D}^0$ events \cite{CLEO1} and $2.4\times10^6$ $D^+D^-$ events \cite{CLEO2} produced in CLEO Collaboration every year. From the Fig.~4, Fig.~5 and Fig.~7, we can find that the maximum branching fractions of $D^+$ and $D^0$ decays approach $10^{-6}$. But if the detection efficiency is considered, the decay modes of $D$ would be difficult to detect.

\begin{table}[htbp]
\caption{Branching Fraction of $B^0\rightarrow
D^-\ell^+\ell^+M_2^-$ and corresponding Reconstruction Events
estimated using Belle's data.} \vspace{0.5em}\centering
\begin{tabular}{ccccccc} \toprule[1.5pt]
&\multicolumn{3}{c}{Branching Fraction}&\multicolumn{3}{c}{Reconstruction Events}\\
\raisebox{1.2ex}[0pt]{$M_2$}
&$e^+e^+$&$e^+\mu^+$&$\mu^+\mu^+$&{$e^+e^+$}&{$e^+\mu^+$}&{$\mu^+\mu^+$}\\
\midrule[1pt]
$\pi$&$10^{-5}\thicksim10^{-6}$&$10^{-5}\thicksim10^{-6}$&$10^{-5}\thicksim10^{-6}$&$4600\thicksim460$&$4600\thicksim460$&$4600\thicksim460$\\
$K$&$10^{-6}\thicksim10^{-7}$&$10^{-6}\thicksim10^{-7}$&$10^{-6}\thicksim10^{-7}$&$420\thicksim42$&$420\thicksim42$&$420\thicksim42$\\
$\rho$&$10^{-4}\thicksim10^{-5}$&$10^{-4}\thicksim10^{-5}$&$10^{-4}\thicksim10^{-5}$&$29700\thicksim2970$&$29700\thicksim2970$&$29700\thicksim2970$\\
$K^*$&$10^{-6}\thicksim10^{-7}$&$10^{-6}\thicksim10^{-7}$&$10^{-6}\thicksim10^{-7}$&$280\thicksim28$&$280\thicksim28$&$280\thicksim28$\\
$D$&$10^{-8}$&$10^{-8}$&$10^{-8}$&4&4&4\\
$D_s$&$10^{-6}\thicksim10^{-7}$&$10^{-6}$&$10^{-6}$&$360\thicksim36$&360&360\\
\bottomrule[1.5pt]
\end{tabular} \label{B0D}
\end{table}
\begin{table}[htbp]
\caption{Branching Fraction of $B^+\rightarrow
\bar{D}^0\ell^+\ell^+M_2^-$ and corresponding Reconstruction
Events estimated using Belle's data.} \vspace{0.5em}\centering
\begin{tabular}{ccccccc} \toprule[1.5pt]
&\multicolumn{3}{c}{Branching Fraction}&\multicolumn{3}{c}{Reconstruction Events }\\
\raisebox{1.2ex}[0pt]{$M_2$}
&$e^+e^+$&$e^+\mu^+$&$\mu^+\mu^+$&{$e^+e^+$}&{$e^+\mu^+$}&{$\mu^+\mu^+$}\\
\midrule[1pt]
$\pi$&$10^{-5}\thicksim10^{-6}$&$10^{-5}\thicksim10^{-6}$&$10^{-5}\thicksim10^{-6}$&$4930\thicksim493$&$4930\thicksim493$&$4930\thicksim493$\\
$K$&$10^{-6}\thicksim10^{-7}$&$10^{-6}\thicksim10^{-7}$&$10^{-6}\thicksim10^{-7}$&$450\thicksim45$&$450\thicksim45$&$450\thicksim45$\\
$\rho$&$10^{-4}\thicksim10^{-5}$&$10^{-4}\thicksim10^{-5}$&$10^{-4}\thicksim10^{-5}$&$31600\thicksim3160$&$31600\thicksim3160$&$31600\thicksim3160$\\
$K^*$&$10^{-6}\thicksim10^{-7}$&$10^{-6}\thicksim10^{-7}$&$10^{-6}\thicksim10^{-7}$&$300\thicksim30$&$300\thicksim30$&$300\thicksim30$\\
$D$&$10^{-7}\thicksim10^{-8}$&$10^{-7}\thicksim10^{-8}$&$10^{-8}$&$40\thicksim4$&$40\thicksim4$&4\\
$D_s$&$10^{-6}$&$10^{-6}$&$10^{-6}\thicksim10^{-7}$&380&380&$380\thicksim38$\\
\bottomrule[1.5pt]
\end{tabular} \label{BD0}
\end{table}
\begin{table}[htbp]
\caption{Branching Fraction of
$B^+{\rightarrow}\pi^0\ell^+\ell^+M_2^-$ and corresponding
Reconstruction Events estimated using Belle's data.}
\vspace{0.5em}\centering
\begin{tabular}{ccccccc} \toprule[1.5pt]
&\multicolumn{3}{c}{Branching Fraction}&\multicolumn{3}{c}{Reconstruction Events}\\
\raisebox{1.2ex}[0pt]{$M_2$}
&$e^+e^+$&$e^+\mu^+$&$\mu^+\mu^+$&{$e^+e^+$}&{$e^+\mu^+$}&{$\mu^+\mu^+$}\\
\midrule[1pt]
$\pi$&$10^{-7}\thicksim10^{-8}$&$10^{-7}\thicksim10^{-8}$&$10^{-7}\thicksim10^{-8}$&$40\thicksim4$&$40\thicksim4$&$40\thicksim4$\\
$\rho$&$10^{-6}\thicksim10^{-7}$&$10^{-6}\thicksim10^{-7}$&$10^{-6}\thicksim10^{-7}$&$250\thicksim25$&$250\thicksim25$&$250\thicksim25$\\
\bottomrule[1.5pt]
\end{tabular} \label{Bpi}
\end{table}

In conclusion, we extended the previous studies to the 4-body LNV (LFV) rare decays of heavy mesons $B$ and $D$, since the 4-body decays share the same vertexes and mixing parameters as well as the CKM matrix elements with the corresponding 3-body decays. Relatively large branching fractions which are comparable with the 3-body decays are obtained, some channels can be reached by current experiments, especially the processes $B\rightarrow {D}\ell^+\ell^+M_2$ when $M_2$ are $\pi$, $K$ and $\rho$.

\section*{Acknowledgments}
We would like to thank Tao Han for his suggestions to carry out this research and providing the FORTRAN codes Hanlib for the calculations. We are also very grateful to Yoshi Sakai for offering the data of particle reconstruction efficiency in Belle Collaboration. This work was supported in part by the National Natural Science Foundation of China (NSFC) under grant No.~11175051.

\begin{figure}[!h]
\centering
\subfigure {}
\addtocounter{subfigure}{-2}
\subfigure[$B^0{\rightarrow}D^-e^+e^+M_2^-$]
{\subfigure{\includegraphics[scale=0.4,trim= 0 0 0 70]{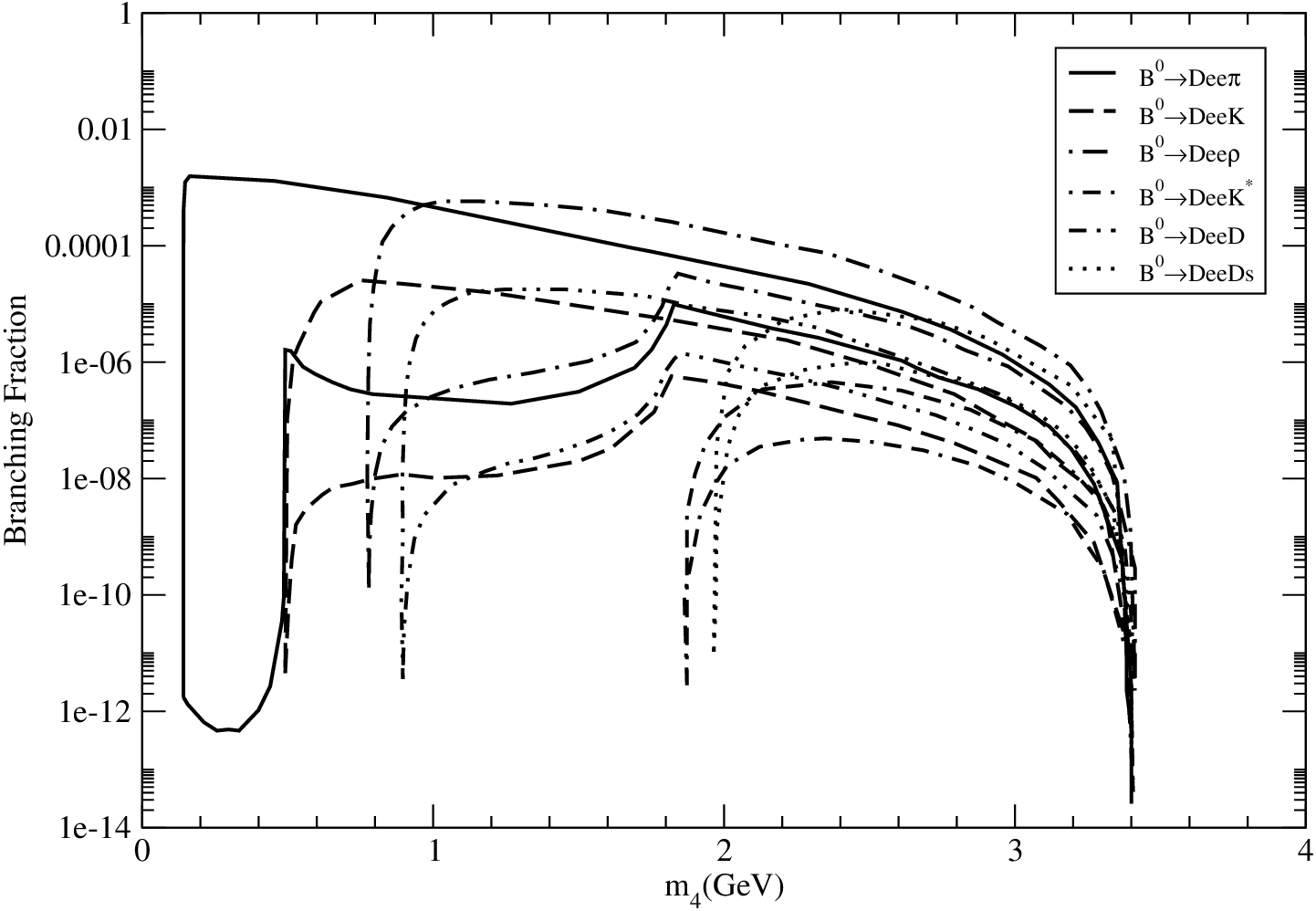}}}
\subfigure {}
\addtocounter{subfigure}{-2}
\subfigure[$B^0{\rightarrow}D^-e^+\mu^+M_2^-$]
{\subfigure{\includegraphics[scale=0.4,trim= 0 0 0 -60]{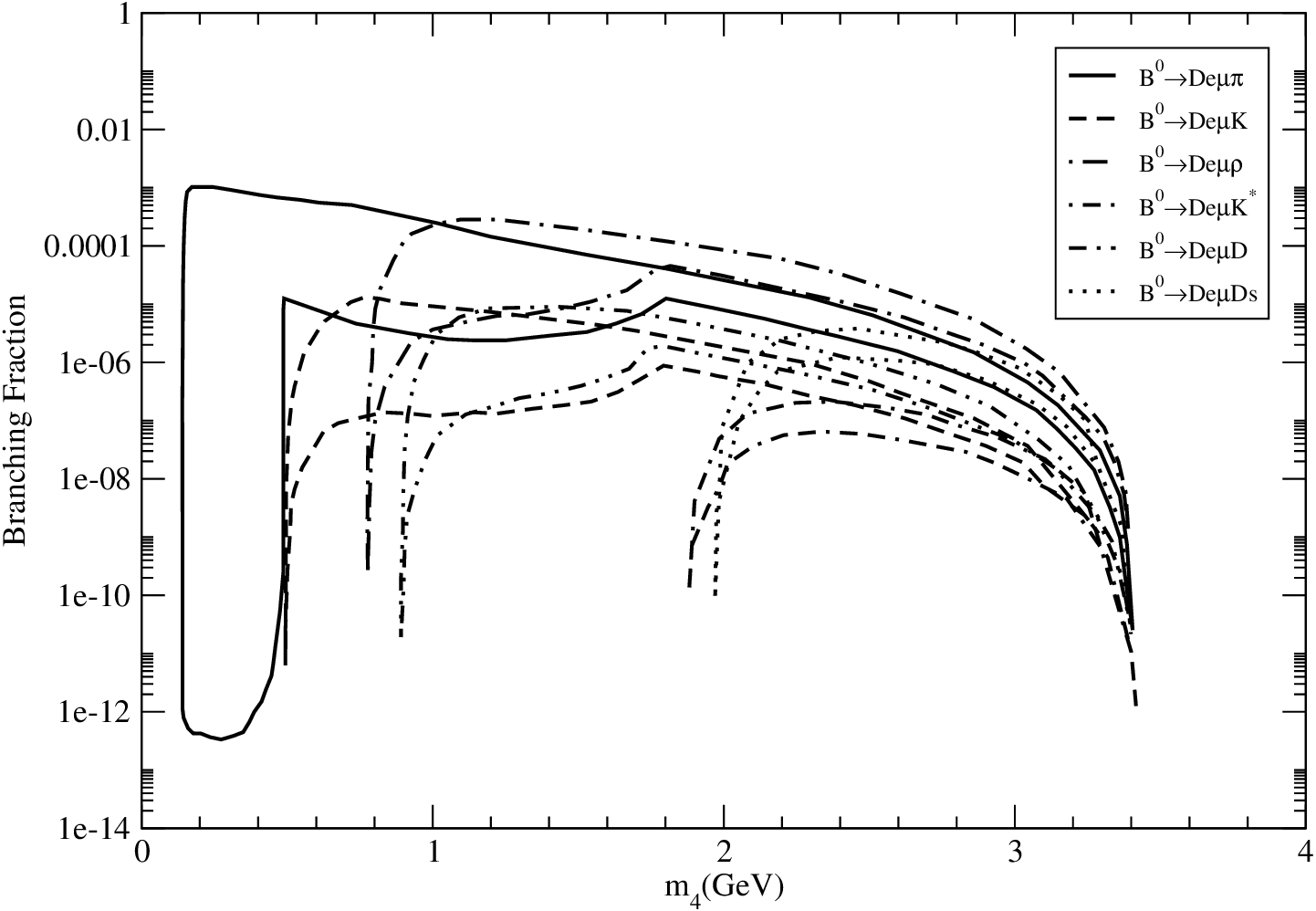}}}
\subfigure {}
\addtocounter{subfigure}{-2}
\subfigure[$B^0{\rightarrow}D^-\mu^+\mu^+M_2^-$]
{\subfigure{\includegraphics[scale=0.4,trim= 0 0 0 -60]{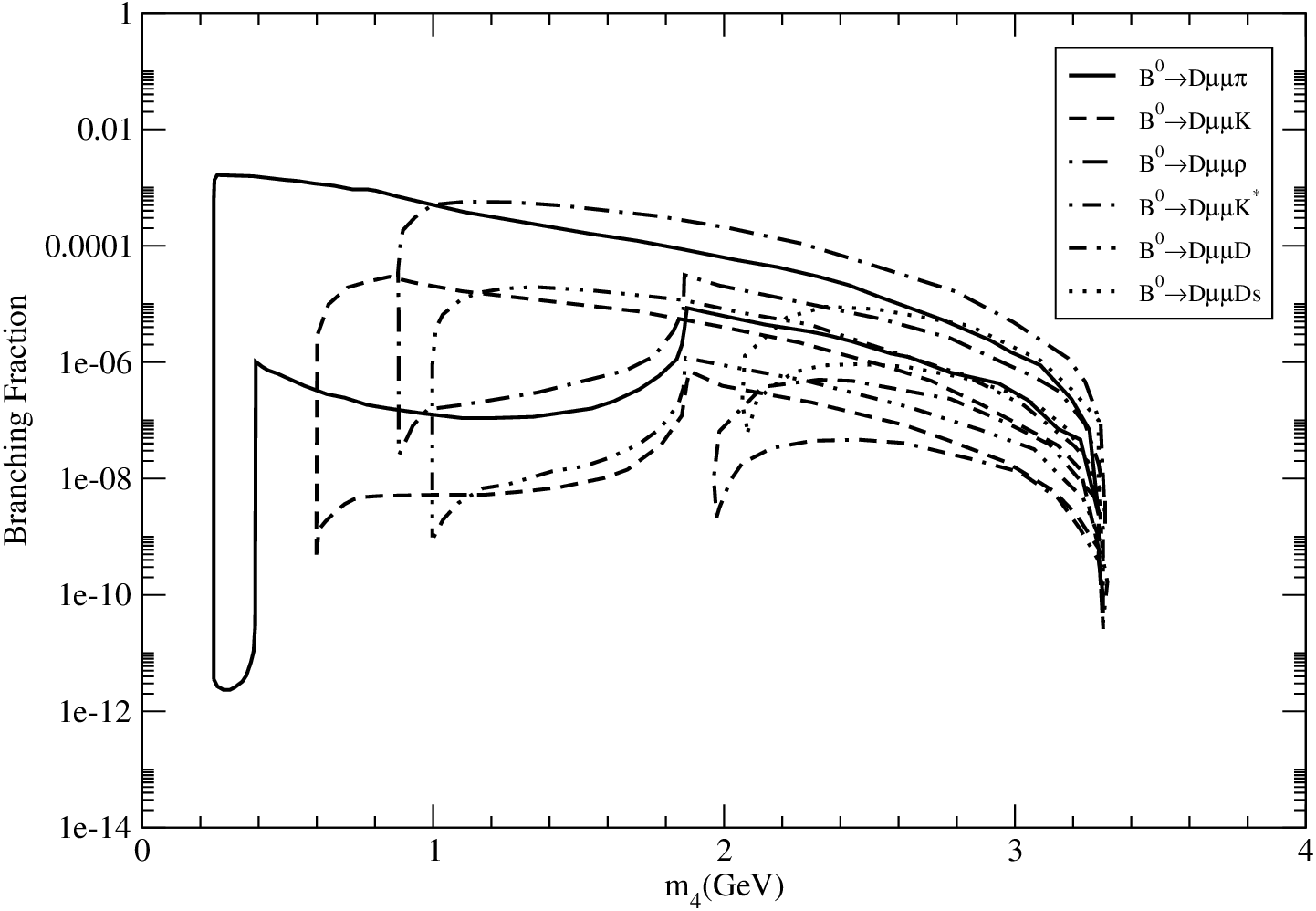}}}
\caption{Theoretically excluded regions inside the curve for the branching fraction of $B^0{\rightarrow}D^-\ell^+\ell^+M_2^-$}\vspace{-1em}
\end{figure}\label{fig:B0}
\begin{figure}[!h]\label{B}
\centering
\subfigure {}
\addtocounter{subfigure}{-2}
\subfigure[ $B^+{\rightarrow}\bar{D}^0e^+e^+M_2^-$]
{\subfigure{\includegraphics[scale=0.4,trim= 0 0 0 70]{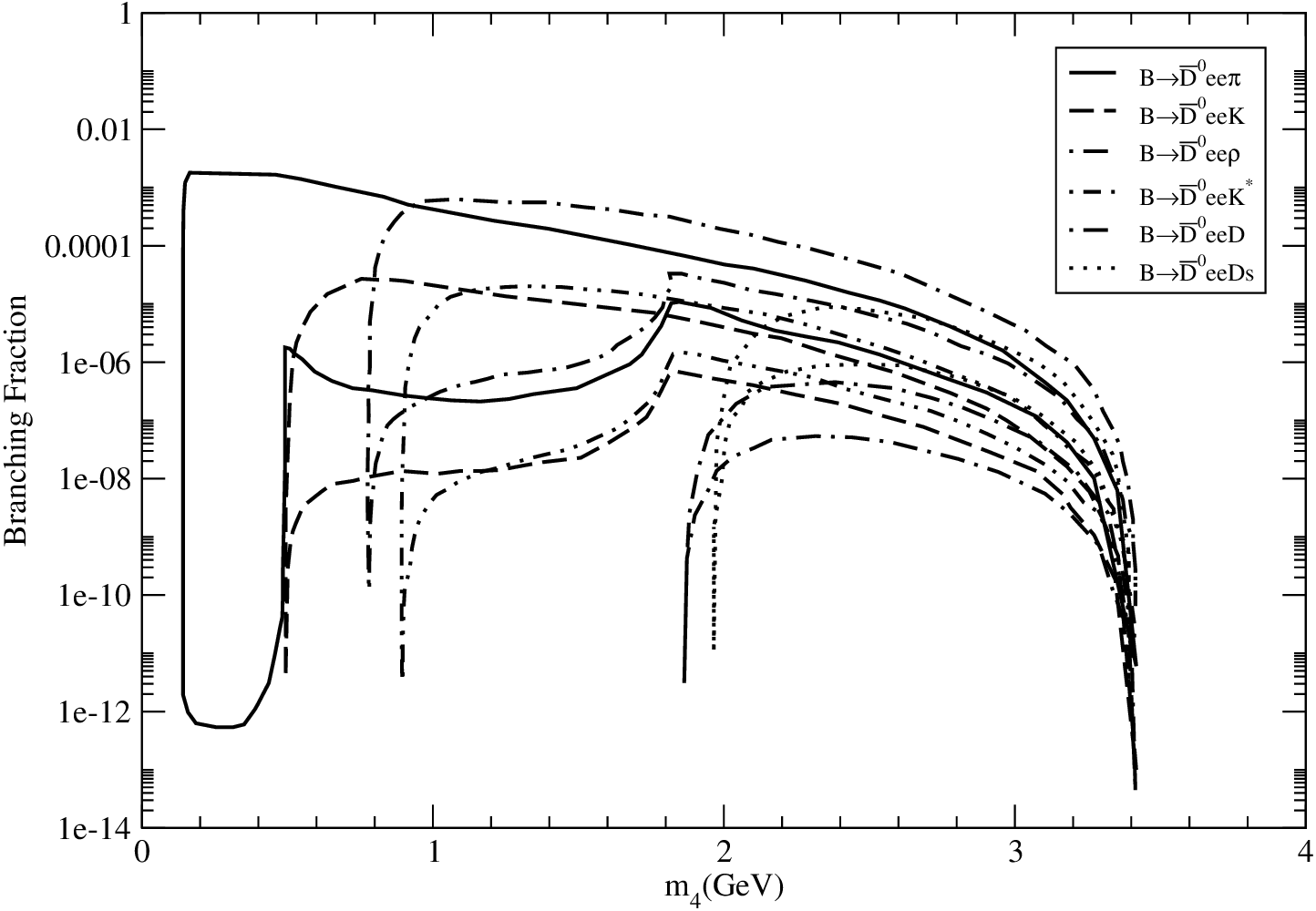}}}
\subfigure {}
\addtocounter{subfigure}{-2}
\subfigure[$B^+{\rightarrow}\bar{D}^0e^+\mu^+M_2^-$]
{\subfigure{\includegraphics[scale=0.4,trim= 0 0 0 -60]{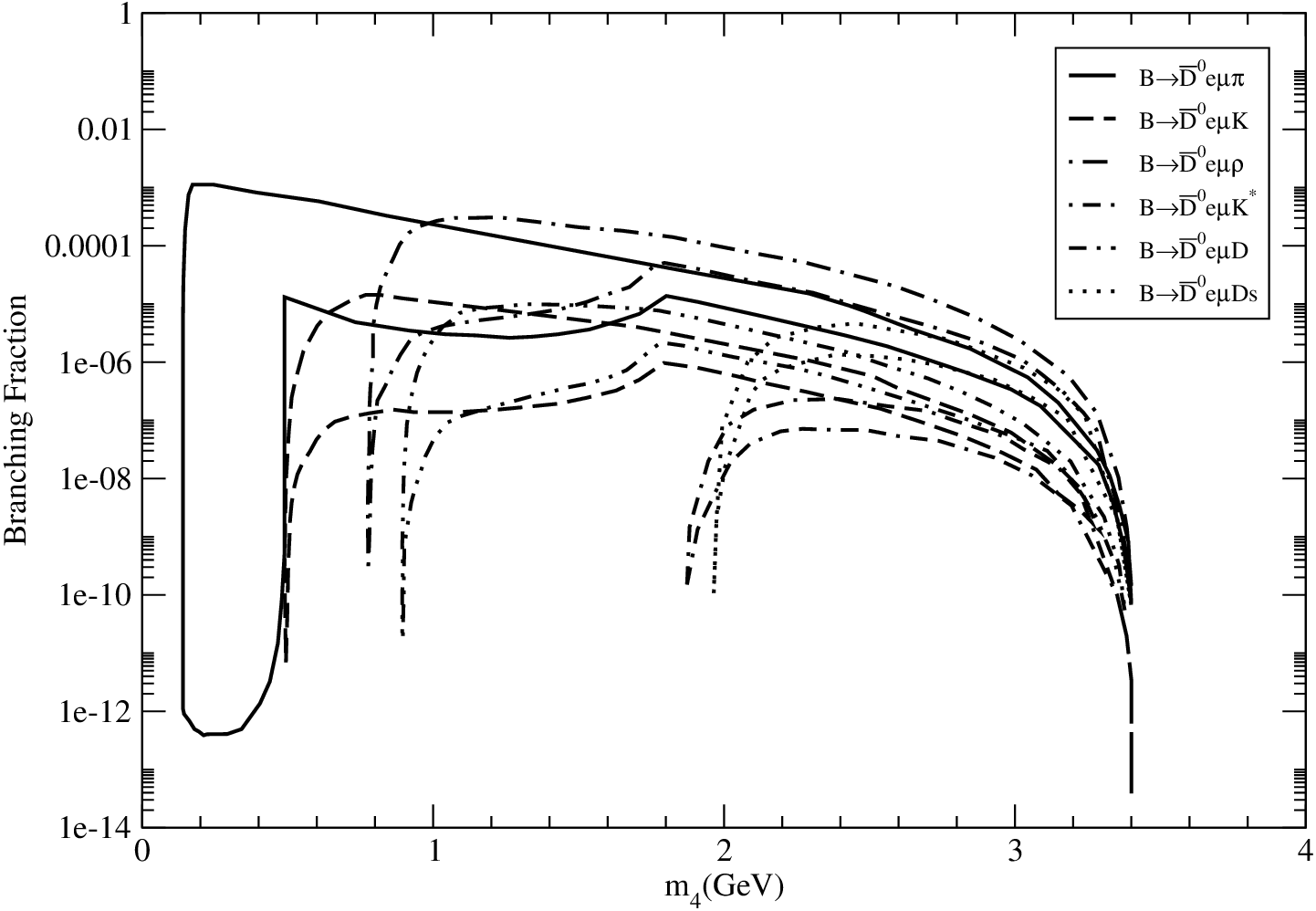}}}
\subfigure {}
\addtocounter{subfigure}{-2}
\subfigure[ $B^+{\rightarrow}\bar{D}^0\mu^+\mu^+M_2^-$]
{\subfigure{\includegraphics[scale=0.4,trim= 0 0 0 -60]{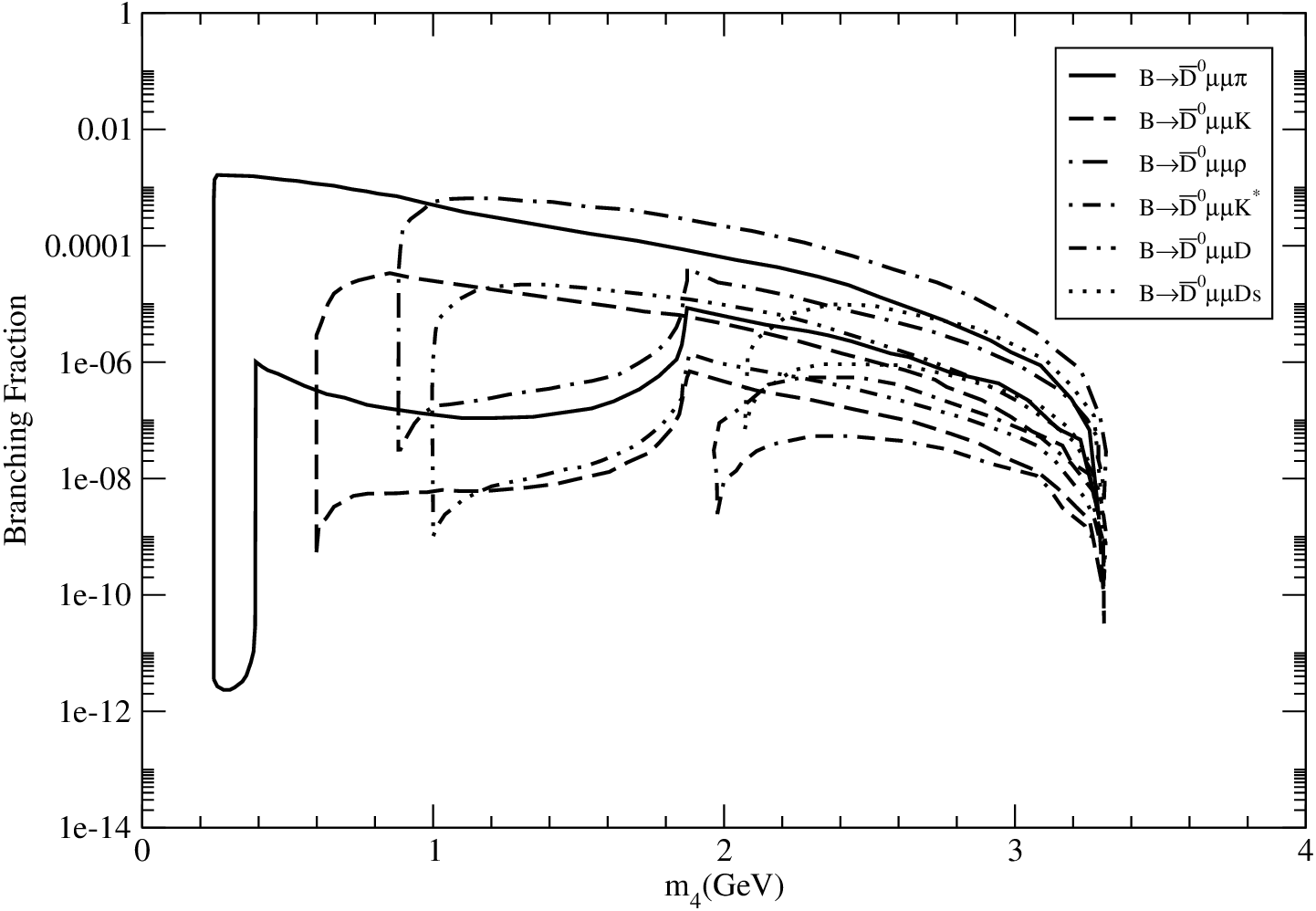}}}
\caption{Theoretically excluded regions inside the curve for the branching fraction of $B^+{\rightarrow}\bar{D}^0\ell^+\ell^+M_2^-$}\vspace{-1em}
\end{figure}
\begin{figure}[!h]\label{D0}
\centering
\subfigure {}
\addtocounter{subfigure}{-2}
\subfigure[ $D^0{\rightarrow}K^-e^+e^+M_2^-$]
{\subfigure{\includegraphics[scale=0.4,trim= 0 0 0 70]{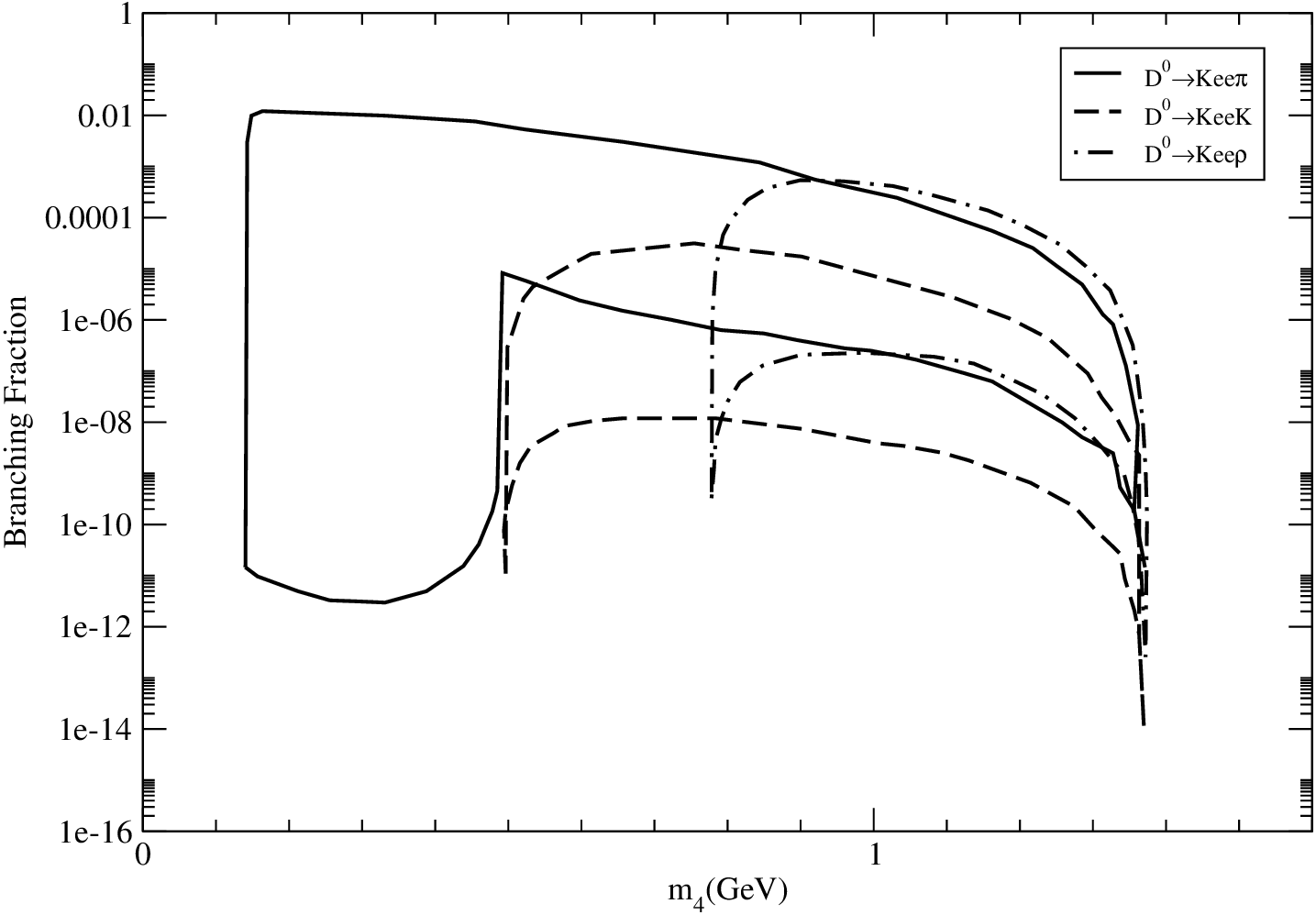}}}
\subfigure {}
\addtocounter{subfigure}{-2}
\subfigure[ $D^0{\rightarrow}K^-e^+\mu^+M_2^-$]
{\subfigure{\includegraphics[scale=0.4,trim= 0 0 0 -60]{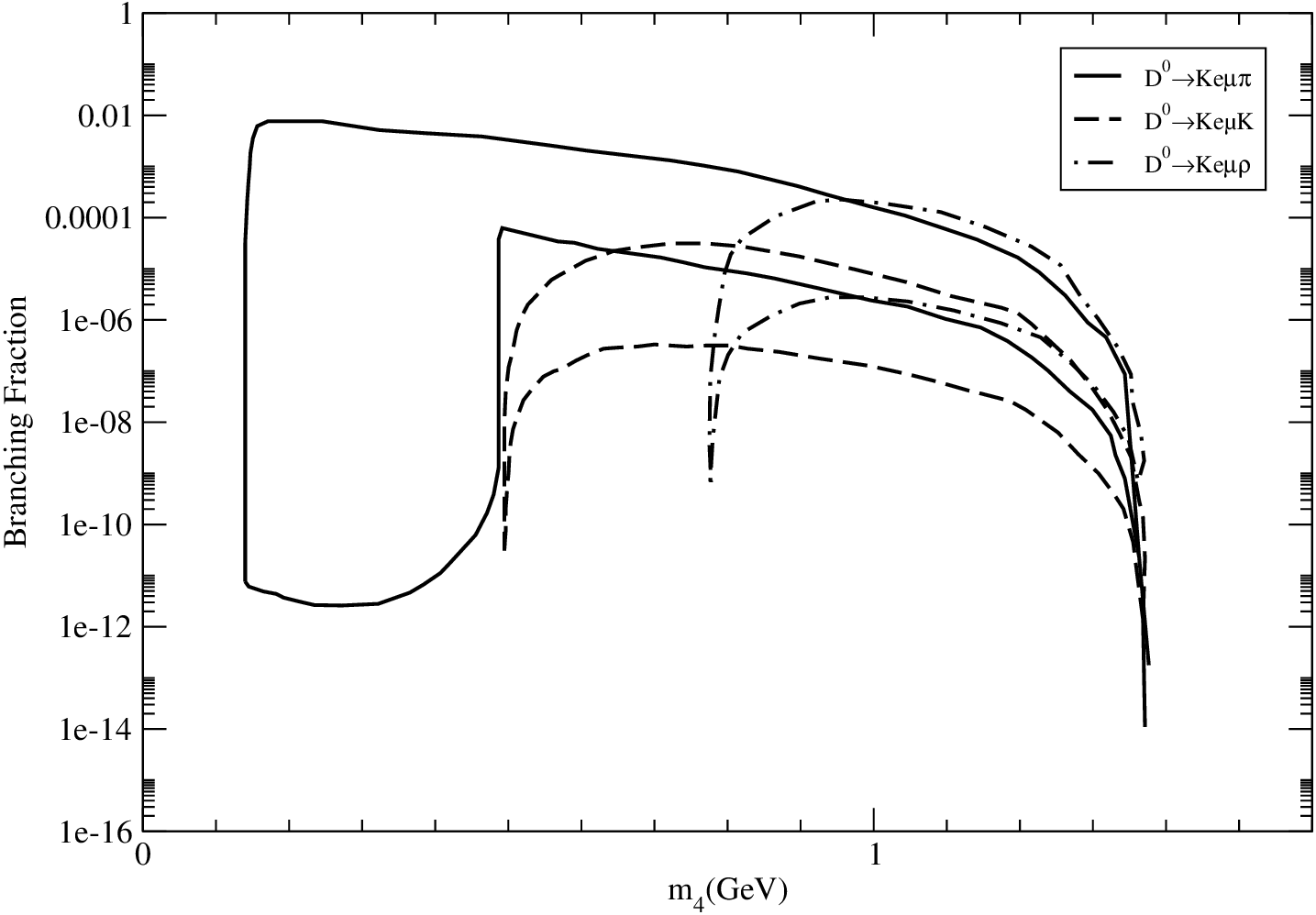}}}
\subfigure {}
\addtocounter{subfigure}{-2}
\subfigure[$D^0{\rightarrow}K^-\mu^+\mu^+M_2^-$]
{\subfigure{\includegraphics[scale=0.4,trim= 0 0 0 -60]{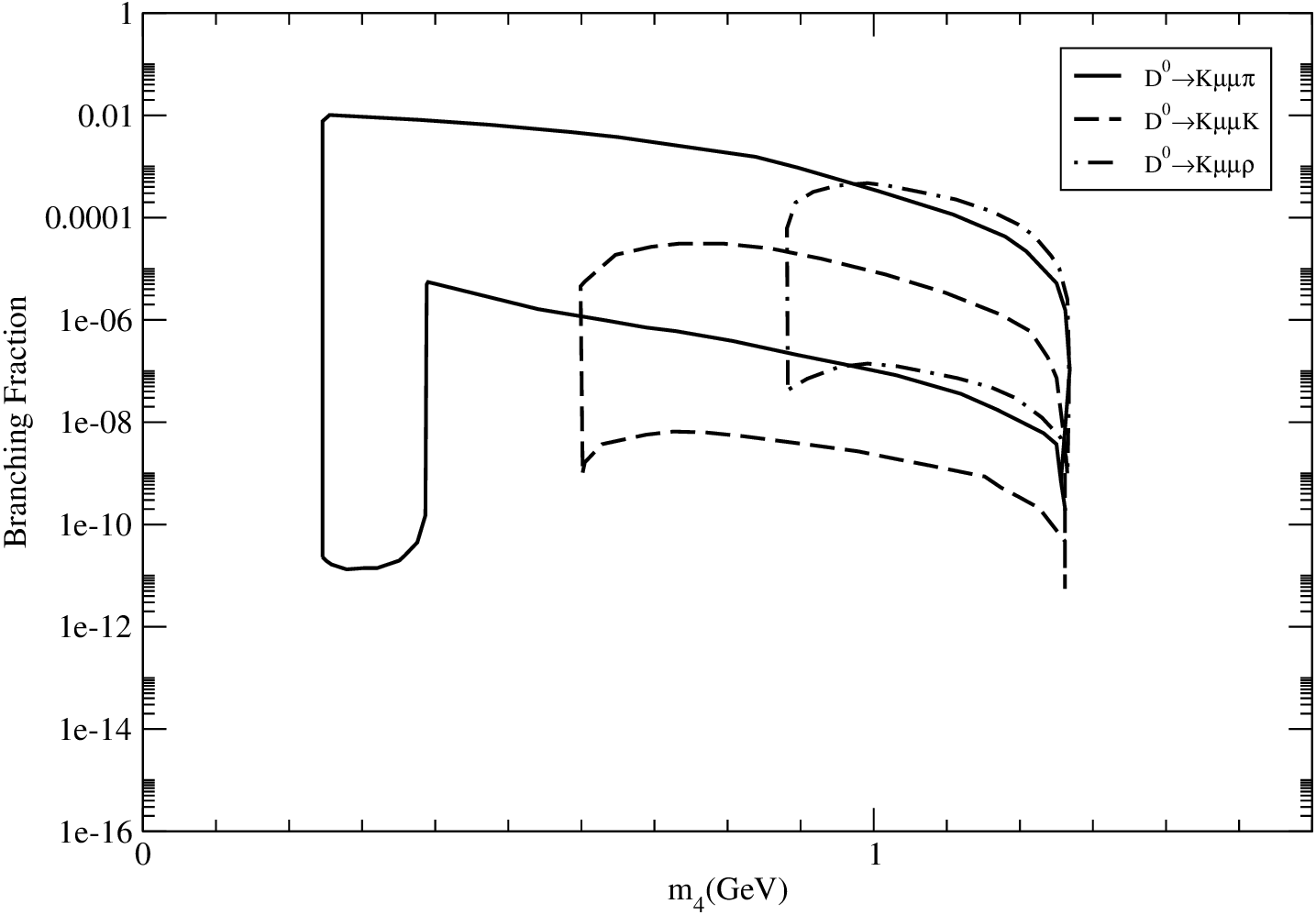}}}
\caption{Theoretically excluded regions inside the curve for the branching fraction of $D^0{\rightarrow}K^-\ell^+\ell^+M_2^-$}\vspace{-1em}
\end{figure}
\begin{figure}[!h]\label{D}
\centering
\subfigure {}
\addtocounter{subfigure}{-2}
\subfigure[$D^+{\rightarrow}\bar{K}^0e^+e^+M_2^-$]
{\subfigure{\includegraphics[scale=0.4,trim= 0 0 0 70]{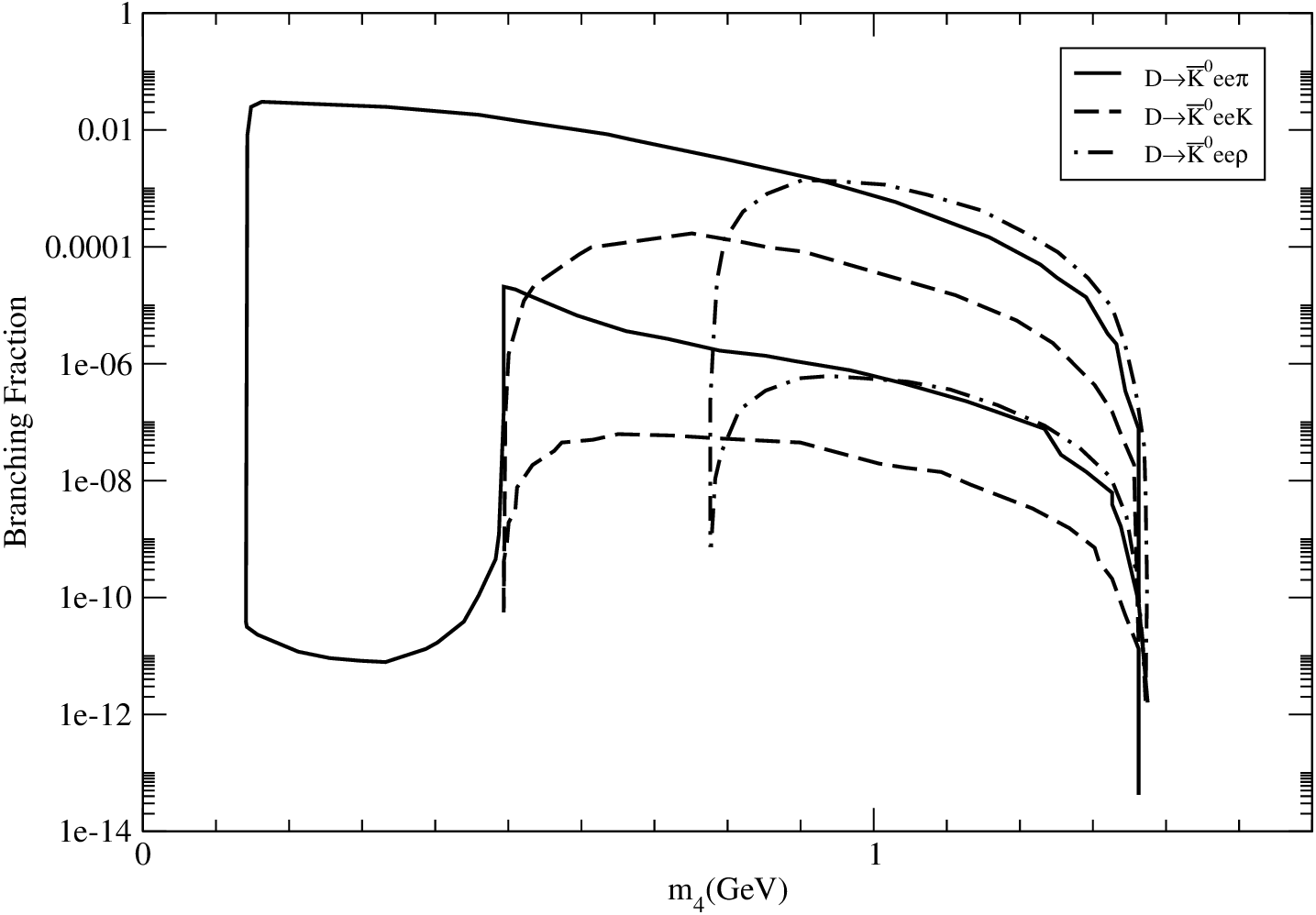}}}
\subfigure {}
\addtocounter{subfigure}{-2}
\subfigure[ $D^+{\rightarrow}\bar{K}^0e^+\mu^+M_2^-$]
{\subfigure{\includegraphics[scale=0.4,trim= 0 0 0 -60]{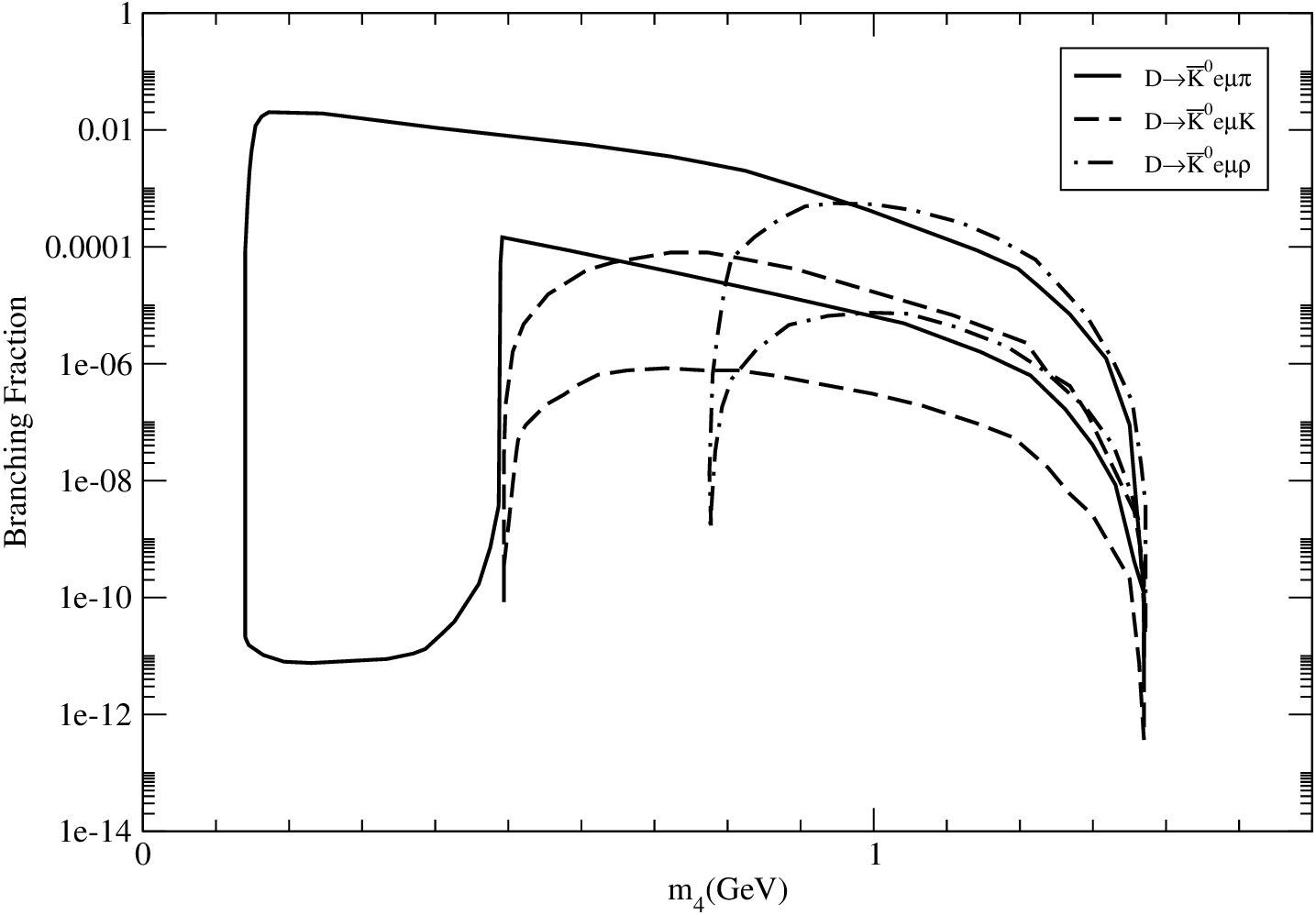}}}
\subfigure {}
\addtocounter{subfigure}{-2}
\subfigure[ $D^+{\rightarrow}\bar{K}^0\mu^+\mu^+M_2^-$]
{\subfigure{\includegraphics[scale=0.4,trim= 0 0 0 -60]{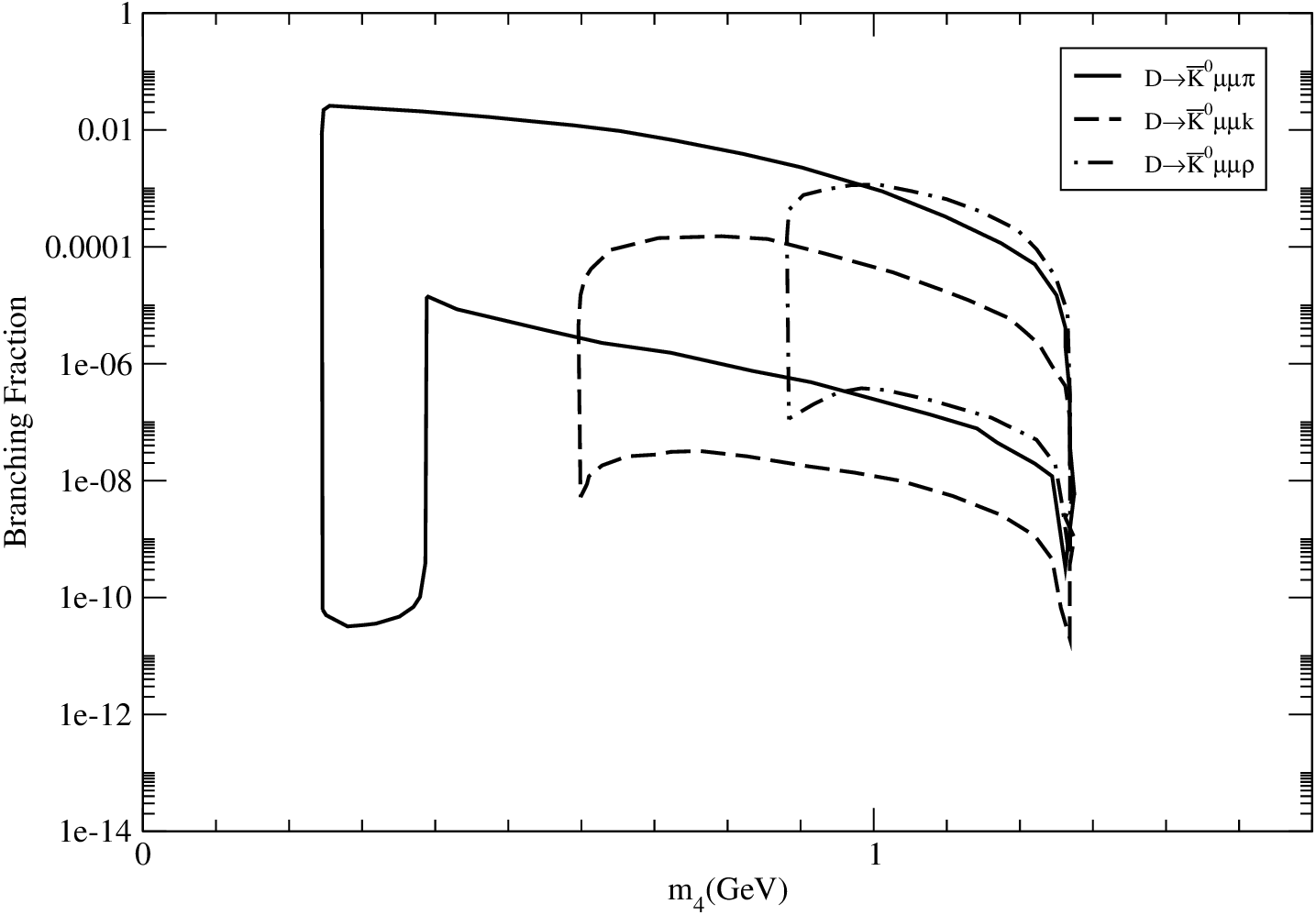}}}
\caption{Theoretically excluded regions inside the curve for the branching fraction of $D^+{\rightarrow}\bar{K}^0\ell^+\ell^+M_2^-$}\vspace{-1em}
\end{figure}
\begin{figure}[!h]\label{Bpi0}
\centering
\subfigure {}
\addtocounter{subfigure}{-2}
\subfigure[ $B^+{\rightarrow}\pi^0e^+e^+M_2^-$]
{\subfigure{\includegraphics[scale=0.4,trim= 0 0 0 70]{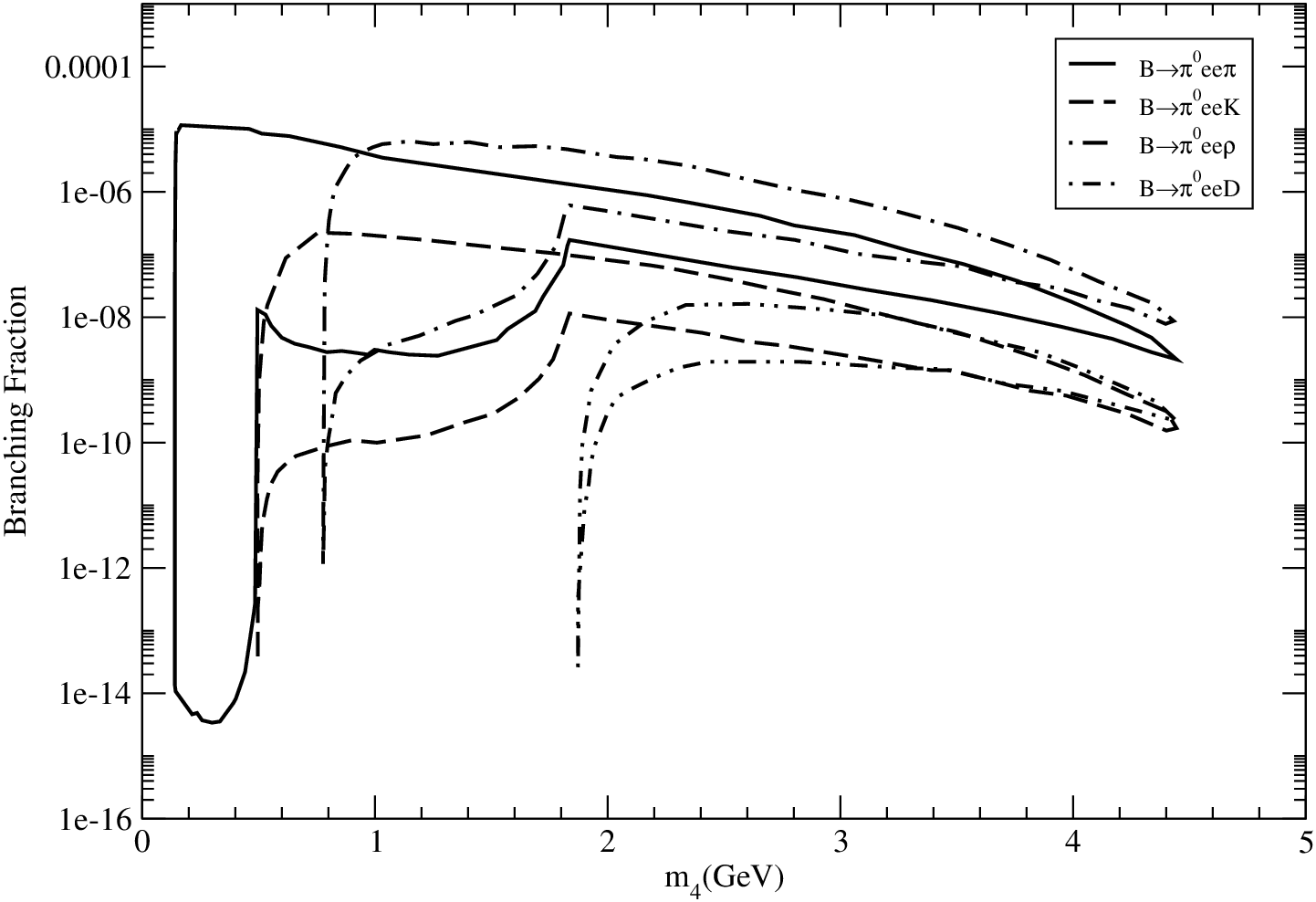}}}
\subfigure {}
\addtocounter{subfigure}{-2}
\subfigure[$B^+{\rightarrow}\pi^0e^+\mu^+M_2^-$]
{\subfigure{\includegraphics[scale=0.4,trim= 0 0 0 -60]{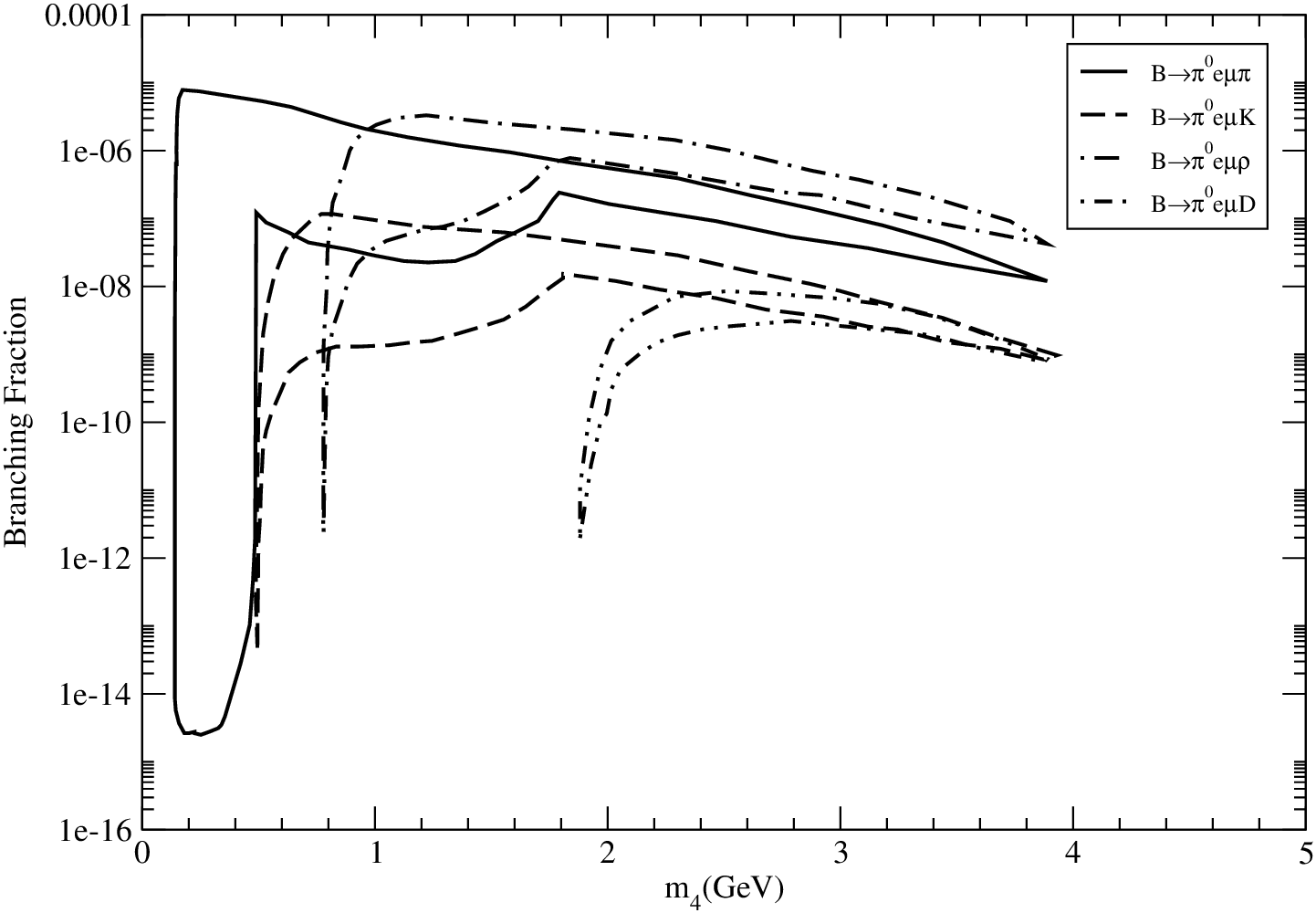}}}
\subfigure {}
\addtocounter{subfigure}{-2}
\subfigure[ $B^+{\rightarrow}\pi^0\mu^+\mu^+M_2^-$]
{\subfigure{\includegraphics[scale=0.4,trim= 0 0 0 -60]{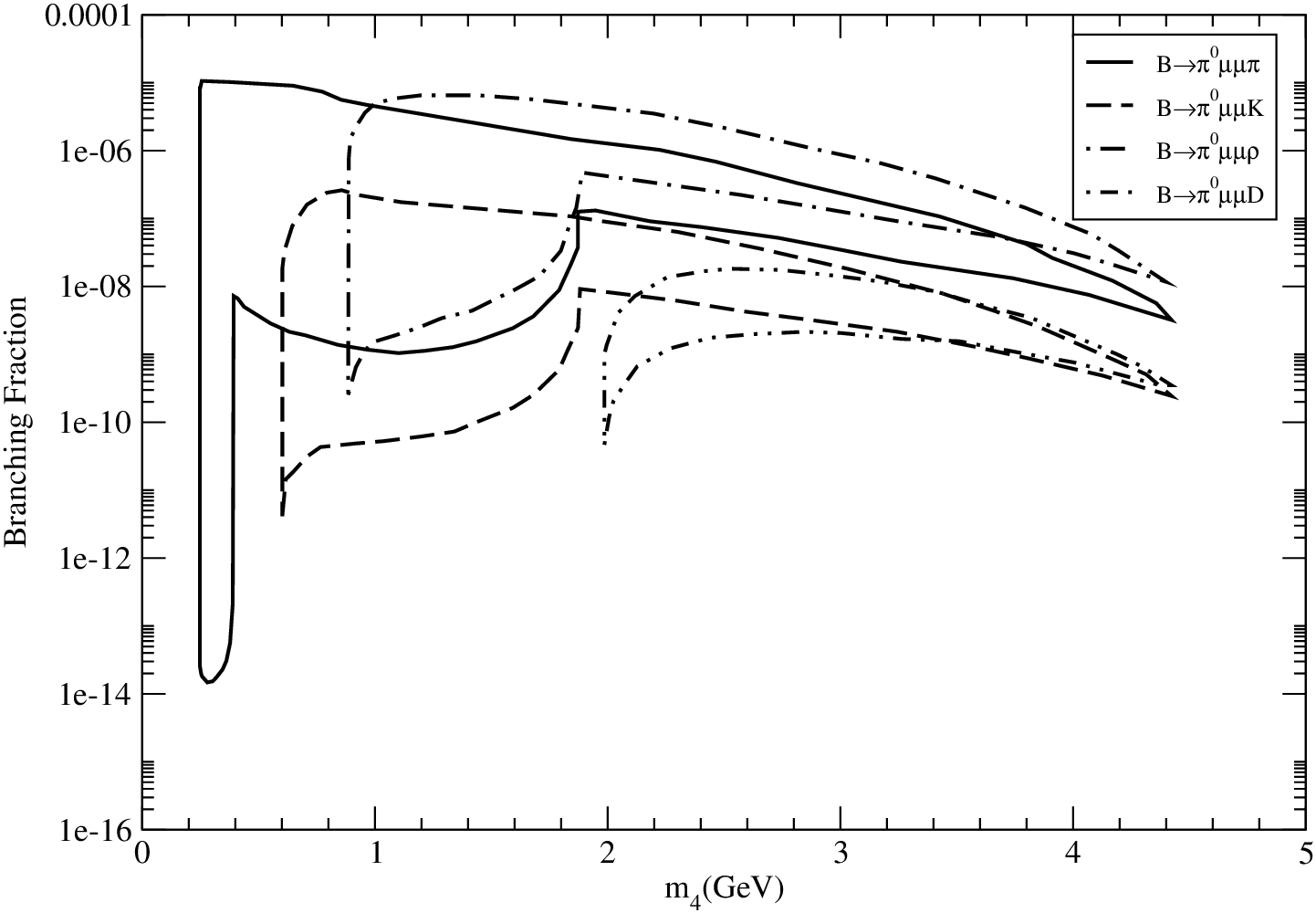}}}
\caption{Theoretically excluded regions inside the curve for the branching fraction of $B^+{\rightarrow}\pi^0\ell^+\ell^+M_2^-$}\vspace{-1em}
\end{figure}
\begin{figure}[!h]\label{Dpi0}
\centering
\subfigure {}
\addtocounter{subfigure}{-2}
\subfigure[$D^+{\rightarrow}\pi^0e^+e^+M_2^-$]
{\subfigure{\includegraphics[scale=0.4,trim= 0 0 0 70]{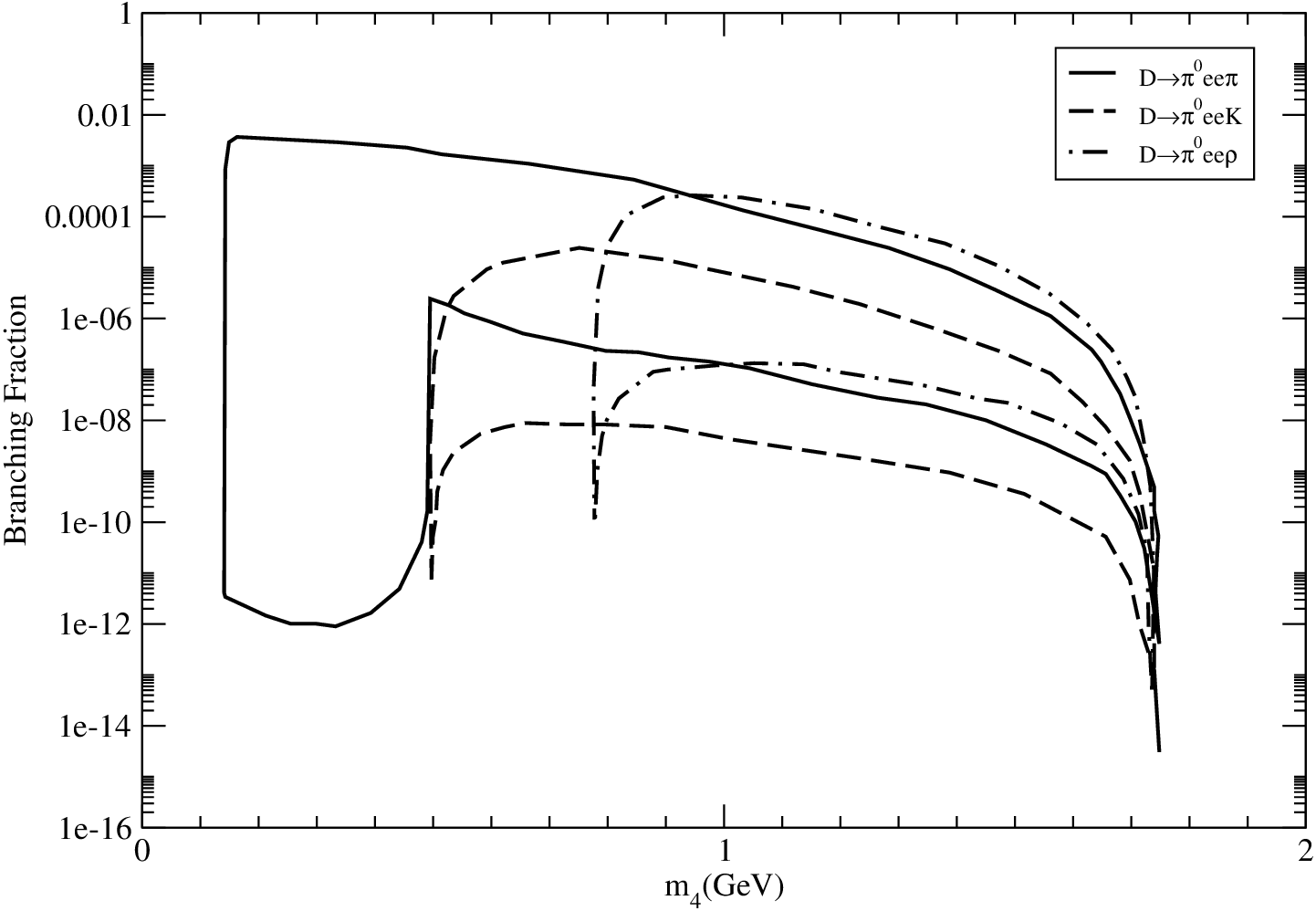}}}
\subfigure {}
\addtocounter{subfigure}{-2}
\subfigure[ $D^+{\rightarrow}\pi^0e^+\mu^+M_2^-$]
{\subfigure{\includegraphics[scale=0.4,trim= 0 0 0 -60]{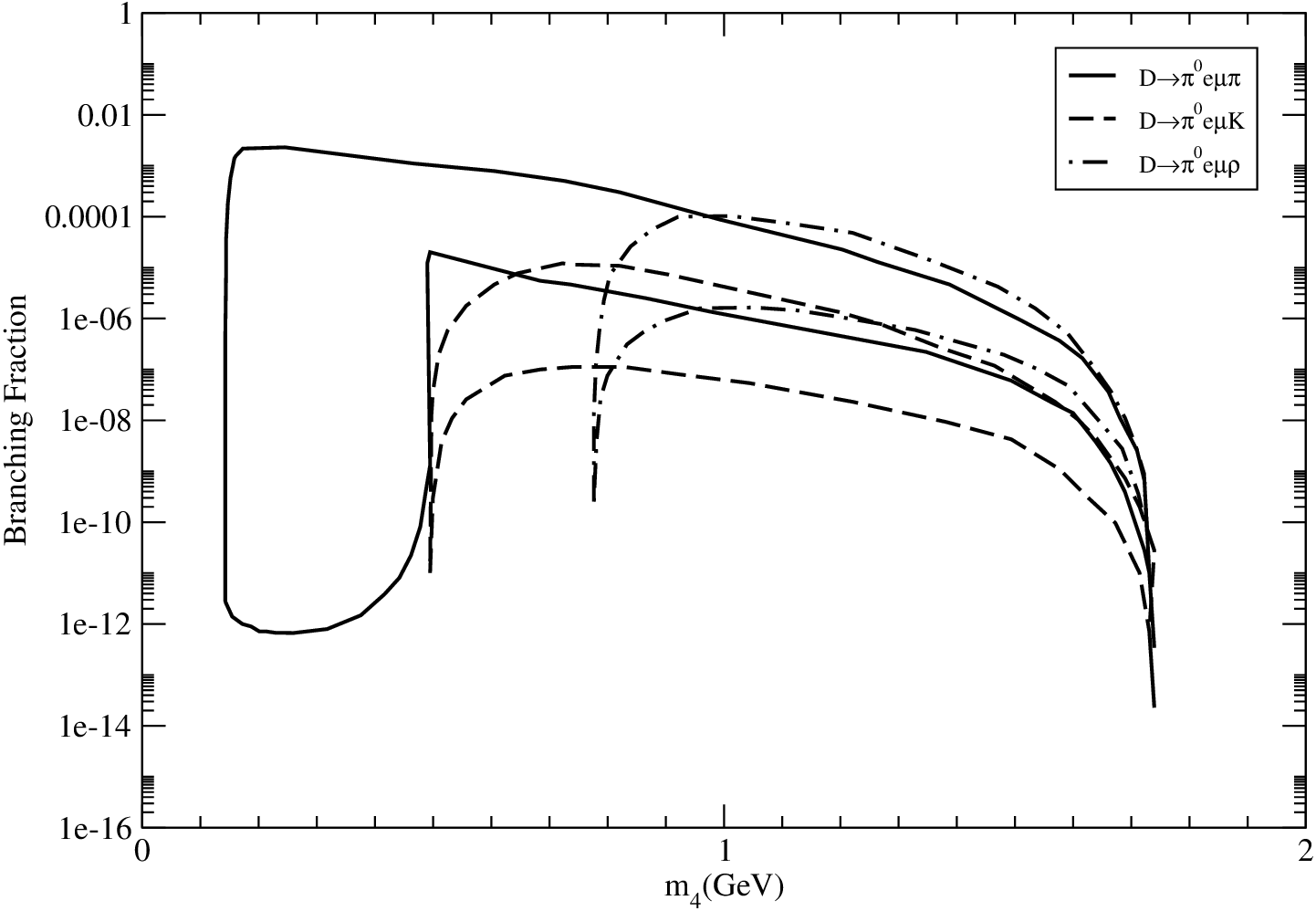}}}
\subfigure {}
\addtocounter{subfigure}{-2}
\subfigure[ $D^+{\rightarrow}\pi^0\mu^+\mu^+M_2^-$]
{\subfigure{\includegraphics[scale=0.4,trim= 0 0 0 -60]{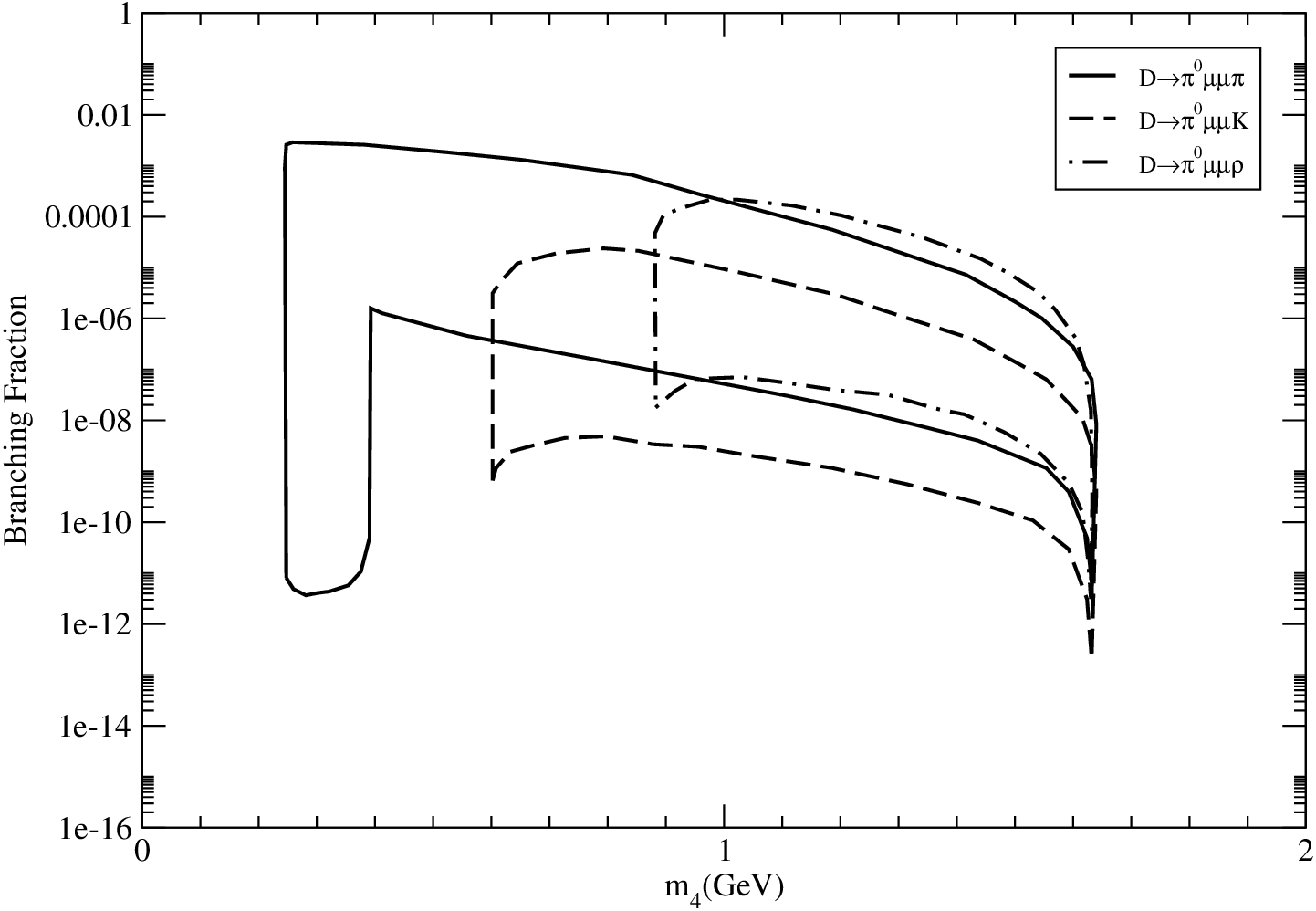}}}
\caption{Theoretically excluded regions inside the curve for the branching frcation of $D^+{\rightarrow}\pi^0\ell^+\ell^+M_2^-$}\vspace{-1em}
\end{figure}
\end{document}